\RequirePackage{fix-cm}
\documentclass[twocolumn]{svjour3}          
\smartqed  
\usepackage[hyphens]{url}
\usepackage{graphicx}
\usepackage{color}
\usepackage[table]{xcolor}
\usepackage{verbatim}
\usepackage{amsmath}
\usepackage{amssymb}
\usepackage{subfig}
\usepackage{url}
\usepackage{bbold}
\usepackage{slashed}
\usepackage{multirow}
\usepackage{threeparttable}
\usepackage{paralist}
\usepackage{bm}
\usepackage{hyperref}
\usepackage{xspace}
\usepackage{rotating}
\usepackage{float,lscape}

\newcommand{\be}{\begin{equation}}
\newcommand{\ee}{\end{equation}}

\newcommand{\pt}{p_{\mathrm{T}}}
\newcommand{\kt}{k_{\mathrm{T}}}
\newcommand{\abseta}{\lvert \eta \rvert}

\newcommand{\ttbar}{\text{t}\overline{\text{t}}}
\newcommand{\bkgrej}{1/\varepsilon_B (\varepsilon_S=30\%)}

\newcommand{\nova}{NOvA}

\newcommand{\sonic}{{\tt{SONIC}}\xspace}
\newcommand{\cmssw}{{\tt{CMSSW}}\xspace}
\newcommand{\externalwork}{{\tt{ExternalWork}}\xspace}
\newcommand{\tensorflow}{{\tt{TensorFlow}}\xspace}
\newcommand{\resnet}{{\tt{ResNet-50}}\xspace}
\newcommand{\protobuf}{{\tt{protobuf}}\xspace}
\newcommand{\gRPC}{{\tt{gRPC}}\xspace}
\newcommand{\remote}{{\it{remote}}\xspace}
\newcommand{\onprem}{{\it{on-prem}}\xspace}
\newcommand{\produce}{{\it{produce}}\xspace}
\newcommand{\acquire}{{\it{acquire}}\xspace}
\newcommand{\delphes}{\textsc{Delphes}}

\journalname{Computing and Software for Big Science}
\def\makeheadbox{{%
\hbox to0pt{\vbox{\baselineskip=10dd\hrule\hbox
to\hsize{\vrule\kern3pt\vbox{\kern3pt
\hbox{\bfseries Computing and Software for Big Science}
\hbox{This is a post-peer-review, pre-copyedit version of this article.}
\hbox{The final authenticated version is available online at: \href{https://doi.org/10.1007/s41781-019-0027-2}{https://doi.org/10.1007/s41781-019-0027-2}.}
\kern3pt}\hfil\kern3pt\vrule}\hrule}%
\hss}}}
\begin{document}

\title{FPGA-accelerated machine learning inference as a service for particle physics computing
\thanks{
J.D., B.H., S.J., B.K., M.L., K.P., N.T., and A.T. are supported by Fermi Research Alliance, LLC under Contract No. DE-AC02-07CH11359 with the U.S. Department of Energy, Office of Science, Office of High Energy Physics.
P.H. and D.R. are supported by a Massachusetts Institute of Technology University grant. 
M.P., J.N., and V.L. received funding from the European Research Council (ERC) under the European Union's Horizon 2020 research and innovation program (grant agreement no 772369).
V.L. also received funding from the Ministry of Education, Science, and Technological Development of the Republic of Serbia under project ON171017.
S-C.H. is supported by DOE Office of Science, Office of High Energy Physics Early Career Research program under Award No. DE-SC0015971. 
S.H., M.T., and D.W. are supported by F5 Networks. 
Z. W. is supported by the National Science Foundation under Grants No. 1606321 and 115164.}
}


\author{
Javier Duarte \and Philip Harris
\and Scott Hauck
\and Burt Holzman
\and Shih-Chieh Hsu
\and Sergo Jindariani
\and Suffian Khan
\and Benjamin Kreis
\and Brian Lee
\and Mia Liu
\and Vladimir Lon\v{c}ar
\and Jennifer Ngadiuba
\and Kevin Pedro
\and Brandon Perez
\and Maurizio Pierini
\and Dylan Rankin
\and Nhan Tran
\and Matthew Trahms
\and Aristeidis Tsaris
\and Colin Versteeg
\and Ted W. Way
\and Dustin Werran
\and Zhenbin Wu
}


\institute{Javier Duarte, Burt Holzman, Sergo Jindariani, Benjamin Kreis, Mia Liu, Kevin Pedro, Nhan Tran, Aristeidis Tsaris \at
              Fermi National Accelerator Laboratory, Batavia, IL 60510, USA
           \and
           Philip Harris, Dylan Rankin \at
              Massachusetts Institute of Technology, Cambridge, MA 02139, USA
           \and
           Scott Hauck, Shih-Chieh Hsu, Matthew Trahms, Dustin Werran \at
              University of Washington, Seattle, WA 98195, USA
           \and
           Suffian Khan, Brian Lee, Brandon Perez, Colin Versteeg, Ted W. Way \at
              Microsoft, Redmond, WA 98052, USA
           \and
           Vladimir Lon\v{c}ar \at
              CERN, CH-1211 Geneva 23, Switzerland\\
              Institute of Physics Belgrade, University of Belgrade, Serbia
           \and
           Jennifer Ngadiuba, Maurizio Pierini \at 
              CERN, CH-1211 Geneva 23, Switzerland
           \and              
           Zhenbin Wu \at 
              University of Illinois at Chicago, Chicago, IL 60607, USA
}         

\date{Received: 29 April 2019 / Accepted: 20 August 2019}

\maketitle

\begin{abstract}
\begin{sloppypar}Large-scale particle physics experiments face challenging demands for high-throughput computing resources both now and in the future.
New heterogeneous computing paradigms on dedicated hardware with increased parallelization, such as Field Programmable Gate Arrays (FPGAs), offer exciting  solutions with large potential gains. 
The growing applications of machine learning algorithms in particle physics for simulation, reconstruction, and analysis are naturally deployed on such platforms.
We demonstrate that the acceleration of machine learning inference as a web service represents a heterogeneous computing solution for particle physics experiments that potentially requires minimal modification to the current computing model.
As examples, we retrain the \resnet convolutional neural network to demonstrate state-of-the-art performance for top quark jet tagging at the LHC and apply a \resnet model with transfer learning for neutrino event classification.
Using Project Brainwave by Microsoft to accelerate the \resnet image classification model,
we achieve average inference times of 60 (10) milliseconds with our experimental physics software framework using Brainwave as a cloud (edge or on-premises) service, representing an improvement by a factor of approximately 30 (175) in model inference latency over traditional CPU inference in current experimental hardware.
A single FPGA service accessed by many CPUs achieves a throughput of 600--700 inferences per second using an image batch of one, comparable to large batch-size GPU throughput and significantly better than small batch-size GPU throughput.
Deployed as an edge or cloud service for the particle physics computing model, coprocessor accelerators can have a higher duty cycle and are potentially much more cost-effective.
\end{sloppypar}
\end{abstract}

\tableofcontents

\section{Introduction}
\label{sec:introduction}
\begin{sloppypar}With large datasets and high data acquisition rates, high-performance and high-throughput computing resources are an essential element of the experimental particle physics program.
These experiments are constantly increasing in both sophistication of detector technology and intensity of particle beams. 
As such, particle physics datasets are growing in size just as the algorithms that process the data are growing in complexity.  
For example, the high luminosity phase of the Large Hadron Collider (HL-LHC) will deliver 15 times more data than the current LHC run.
The HL-LHC will collide bunches of protons at a rate of 40~MHz, and the collision environment will have 5 times as many particles per collision~\cite{Apollinari:2116337}.
The Compact Muon Solenoid (CMS) experiment will be upgraded for the HL-LHC with up to 10 times more readout channels. 
Through a series of online filters, CMS aims to store HL-LHC collision events at a rate of 5~kHz.
Such a data rate leads to datasets that are exabytes in scale~\cite{Alves:2017she}.  
Future neutrino experiments such as Deep Underground Neutrino Experiment (DUNE)~\cite{Acciarri:2016crz} and cosmology experiments like Square Kilometre Array~(SKA)~\cite{mellema2013g} are expected to produce datasets at the exabyte scale.
\end{sloppypar}

In the past, the physics and computing communities relied largely on the progress of silicon technologies to handle growing computing requirements.
However, at present, improvement in single processor performance is stalling due to changes in the scaling of power consumption~\cite{NAP12980}.
The current particle physics computing paradigms will not suffice to simulate, process, and analyze the massive datasets that the next-generation experimental facilities will deliver.
New technologies that provide order-of-magnitude improvements are needed.

\begin{sloppypar}Concurrently, the ubiquity of sophisticated detectors with complex outputs has led to the quick adoption of machine learning (ML) algorithms as tools to reconstruct physics processes.  
Neutrino experiments currently use state-of-the-art convolutional neural networks (CNNs)~\cite{Acciarri:2016ryt,Aurisano:2016jvx}, such as {\tt{GoogLeNet}} and \resnet~\cite{DBLP:journals/corr/HeZRS15}, to perform the neutrino event reconstruction and identification. 
At the LHC, ML methods are used in all stages of the ATLAS, CMS, LHCb, and ALICE experiments, from low-level calibration of individual reconstructed particles~\cite{Chatrchyan:2013dga} to high-level optimization of final-state event topologies~\cite{Nguyen:2018ugw}.  
ML was a vital component of the Higgs boson discovery~\cite{Chatrchyan:2012xdj,Aad:2012tfa} and is now being explored for the first level of processing: low latency, sub-microsecond online filtering applications~\cite{Duarte:2018ite,Low:2289251}.  
Across big science, such as cosmology and large astrophysical surveys, similar trends exist as the experiments grow and the data rates increase. 
\end{sloppypar}

While the computing challenge in particle physics is a vital concern for current and future experiments, it is not unique.
With the rise of so-called ``big data,'' Internet of Things (IoT), and the increase in the quantity of data across a wide range of scientific fields, the sophisticated large-scale processing of big data has become a global challenge.  
At the forefront of this trend is the need for new computing resources to handle both the training and inference of large ML models.

\begin{sloppypar}In this paper, we focus on the {\bf inference} of deep ML models as a solution for processing large datasets; inference is computationally intensive and runs repeatedly on hundreds
of billions of events.  
A growing trend to improve computing power has been the development of hardware that is dedicated to accelerating certain kinds of computations.
Pairing a specialized coprocessor with a traditional CPU, referred to as {\it heterogeneous computing}, greatly improves performance.
These specialized coprocessors, including GPUs, Field Programmable Gate Arrays (FPGAs), and Application Specific Integrated Circuits (ASICs), utilize natural parallelization and provide higher data throughput.  
ML algorithms, and in particular deep neural networks, are at the forefront of this computing revolution due to their high parallelizability and common computational needs.
\end{sloppypar}

To capitalize on this new wave of heterogeneous computing and specialized hardware, particle physicists have two primary options: 
\begin{enumerate}
\item {\it Adapt domain-specific algorithms to run on specialized accelerator hardware.} \\
This option takes advantage of specific human expert knowledge, but can be challenging to implement on new and ever-changing hardware platforms with different computing paradigms (such as {\tt{CUDA}} or Verilog).  New portable development environments (e.g. {\tt OpenCL},{\tt Kokkos}) can potentially provide cross-hardware solutions.
\item {\it Design ML algorithms to replace domain-specific algorithms.}\\
This option has the advantage of running natively on specialized hardware using open-source software stacks, but it can be a challenge to map specific physics problems onto ML solutions.
\end{enumerate}

In this paper, we explore how such heterogeneous computing resources can be deployed within the current computing model for particle physics in a scalable and non-disruptive way.
While accelerating domain-specific algorithms on specialized hardware is possible, in this paper we study the second option, where a ML algorithm is adapted to solve a challenge and accelerated using a specialized hardware platform.
We will present physics results for a publicly available top quark tagging dataset for the LHC~\cite{topdataset} and discuss how this could be applied for neutrino experiments such as {\nova}~\cite{Ayres:2007tu}.
This study focuses on the newly available Microsoft Project Brainwave platform that deploys FPGA coprocessors as a service at datacenter scale~\cite{configurable-cloud-acceleration}.
Brainwave provides a first scalable platform to study, though other such options exist. 
Results from this study will serve as a performance benchmark for any similar systems and will provide valuable lessons for applying new technologies to particle physics computing.

The rest of this paper is organized as follows.
In Section~\ref{sec:motivation}, we describe the requirements of the particle physics computing model that is used in collider experiments at the LHC and neutrino experiments such as DUNE.  We detail the challenges facing this computing model in the future.  
In Section~\ref{sec:physics}, we explore some example use cases to be deployed on the Microsoft Brainwave platform. We train and evaluate a dedicated model identifying particles at the LHC and discuss the potential application for neutrino physics.
In Section~\ref{sec:sonic}, we then describe the Microsoft Brainwave platform and how we integrate it into our experimental computing model to accelerate ML inference. 
In Section~\ref{sec:results}, we present latency results from tests of FPGA coprocessors as a service and compare the results to benchmark values for CPUs and GPUs.  We also provide first studies on the scalability of such an approach.    
Finally, in Section~\ref{sec:outlook}, we conclude by summarizing the study and discussing the next steps required for further development of this program.

\section{Computing in particle physics}
\label{sec:motivation}
\subsection{Particle physics computing model}

The computing model for many large scale physics experiments is based on processing events.  
An event here is defined as a measurement of some physical process of interest; in the case of the LHC, it is a collision of bunches of protons every 25~ns.
The event consists of complex detector signals that are filtered, combined, and analyzed; typically, the raw signal inputs are converted into objects with a more physical meaning.  
There is both online processing, in which the event is selected from a buffer and analyzed in real time, and offline processing, in which the event has been written to disk and is more thoroughly analyzed with less stringent latency requirements.  
The online processing system, called the \emph{trigger}, reduces the rate of events to a manageable level to be recorded for offline processing.
The trigger is typically divided into multiple tiers. 
The first tier (Level-1, L1) is performed with custom electronics with very low latency (1--10~$\mu\textrm{s}$) where the latency is a fixed size for every event.
The second step (high level trigger, HLT) is performed on more standard computing resources and has a variable per-event latency of 10--100~ms.
Finally, offline analysis of the saved events passing the HLT can take significantly longer, though ultimately the offline processing time is limited by available computing resources.

\begin{figure*}[tbh!]
\begin{center}
\includegraphics[width=0.80\linewidth]{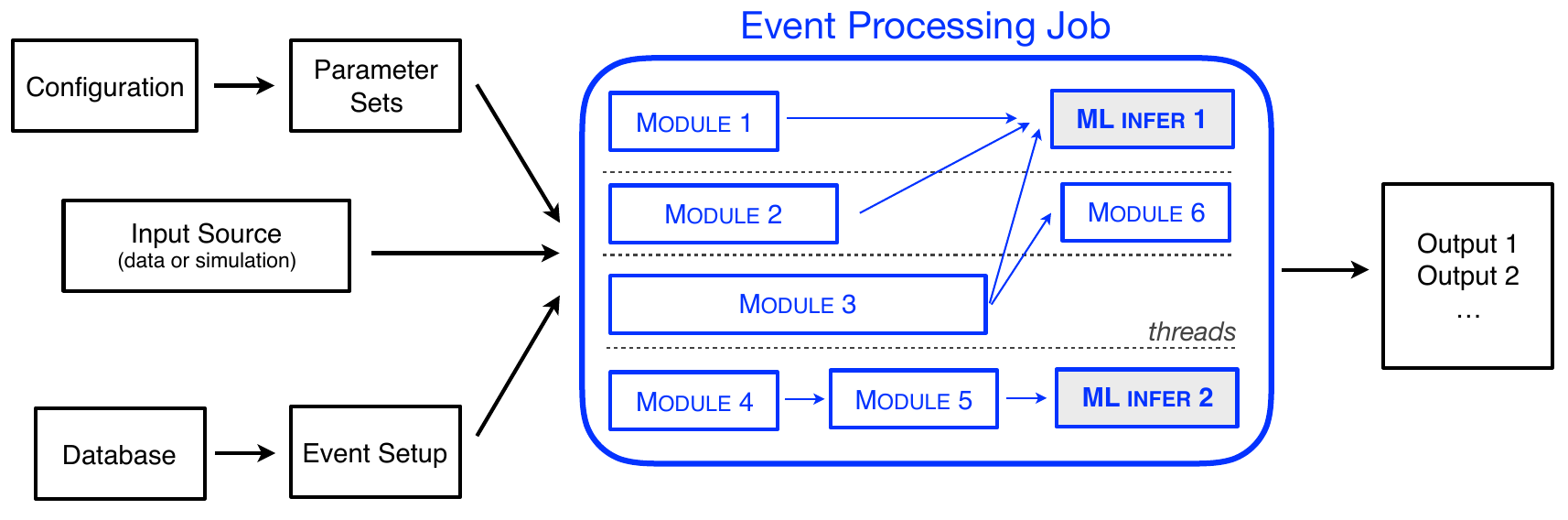}
\end{center}
\caption{A diagram of the computing model used in the CMS software.}
\label{fig:edm}
\end{figure*}

In this paper, we consider the possible gains from heterogeneous computing resources as applied to both the HLT and offline processing steps.
When considering how best to use new optimized computing resources for physics, we must understand the implications of the event processing model described above.
An example of the current computing model is shown in Fig.~\ref{fig:edm}. 
Event data is processed, often sequentially, across multiple CPU threads.

It is important to note that the basic processing unit is a single event and performing the same task for multiple events (batching) becomes significantly more complex to manage.
Because each event contains potentially millions of channels of information, it is optimal to load the needed components of that event into memory and then execute all desired algorithms for that event.
The tasks themselves, denoted in Fig.~\ref{fig:edm} as modules, can be very complex, either with time-consuming physics-based algorithms, or, as is becoming more popular, machine learning algorithms.
There may be dozens or even hundreds of modules executed for each event.
It can be seen that the most time-consuming and complex tasks will be the latency bottleneck in event processing.

\subsection{Upcoming computing challenges}

\begin{sloppypar}In the next decade, the HL-LHC upgrade will increase the LHC collision rate by an order of magnitude.
The CMS detector will undergo a series of upgrades to be able to cope with the increased collision rate and the associated increase in radiation levels, which would damage parts of the current detector beyond the point of recovery.
The detector upgrades include a new pixel tracker with almost 2 billion readout channels and a high granularity endcap calorimeter with 6 million channels~\cite{CMSCollaboration:2015zni}.
Both of these constitute more than an order-of-magnitude increase in channels compared to the current systems.
Another consequence of the HL-LHC upgrade will be an increase in the rate of multiple collisions per proton bunch crossing (pileup).
While the current LHC configuration results in about 30 collisions per bunch crossing, this value will increase to about 200 collisions at the HL-LHC.
\end{sloppypar}

\begin{figure}[!htb]
\begin{center}
\includegraphics[width=0.8\linewidth]{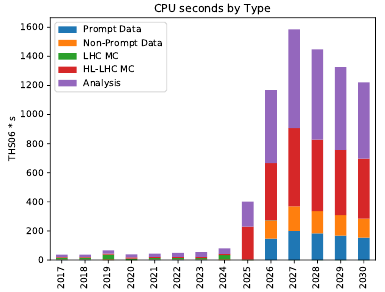}
\end{center}
\caption{Estimated CPU resource needs for CMS in the next decade~\cite{Alves:2017she}. “THS06” stands for tera ($10^{12}$) HEP-SPEC06, a standard measure of the performance of a CPU code used in high-energy physics.}
\label{fig:cpuhllhc}
\end{figure}

The consequence is that the upgraded CMS detector will have to record and process more events, each of which contain more channels and more energy deposits from pileup.
The time to analyze these extremely complex events is currently simulated to be approximately 300~seconds.
The impact on the CPU resources needed by CMS is depicted in Fig.~\ref{fig:cpuhllhc}~\cite{Alves:2017she}.
The relative increase in computing resources required for the HL-LHC is more than a factor of 10 greater than current needs.
Similarly, the DUNE experiment, the largest liquid argon neutrino detector ever designed, will comprise roughly 1 million channels with megahertz sampling and millisecond integration times~\cite{Acciarri:2016crz}.
Both of these frontier experiments will need new solutions for event processing to be able to make sense of the large datasets that will be delivered in the next decades.

\section{Machine learning for physics}
\label{sec:physics}
In this section, we highlight examples of machine learning models relevant for physics to test in accelerator hardware.  
These are not meant as realistic examples, but rather as a proof-of-concept to be expanded when more mature physics models can be accelerated on coprocessors.    

\subsection{\texorpdfstring{\resnet}{ResNet-50} and other models}
\label{sec:resnet}

At the moment, only a limited number of neural network architectures are available for acceleration on the Brainwave platform.  
The available models---\resnet, {\tt VGG-16}~\cite{DBLP:journals/corr/SimonyanZ14a}, and {\tt DenseNet-121}~\cite{DBLP:journals/corr/HuangLW16a}---are CNNs optimized for image classification.  
These CNNs typically contain several convolutional layers that extract meaningful features of the image. 
This part of the network is the most computationally intensive and is often called the ``featurizer.''  
The final part of the network is much smaller and typically includes a few fully connected layers with the final output corresponding to a set of probabilities for each category.
This part of the network is called the ``classifier.''
In our study, we focus on the \resnet model.
The FPGA is used to accelerate the featurizer step of the \resnet inference, while the classifier step is performed on the CPU.
In total, \resnet contains approximately 25 million parameters and requires approximately 4 G-ops ($4 \times 10^9$) for a single inference.  
While the neural network architectures are fixed, the weights can be retrained within one of these available network architectures.  We use this workflow to train a \resnet neural network for a physics-specific task in Sec.~\ref{sec:lhc} and Sec.~\ref{sec:nova}. 
Even with a restricted architecture, the amount of ML tasks that can be performed with these sophisticated image recognition models is substantial.
We will explore two: classification of boosted top quarks and neutrino flavor classification.

\begin{sloppypar}However, we also stress that this is a proof-of-concept study to demonstrate the improvements for physics computing from heterogeneous computing platforms as a service.  As the technology matures rapidly, we will also see an improvement in the software toolsets associated with this new hardware.  We expect the capability to translate any model to specialized hardware to become available in the near future. In fact, several tools are working towards this capability~\cite{XilinxMLSuite,TFforTPU,OpenVino}.
\end{sloppypar}

\subsection{Top tagging at the LHC}
\label{sec:lhc}
\label{sec:results-trainLHC}
At the LHC, quarks and gluons originating from the proton collisions produce collimated sprays of particles in the detector called \emph{jets}. 
Studying the substructure of these jets is an important tool for identifying their origin.
There are broad physics applications from studying Higgs boson properties, to searching for new physics beyond the standard model such as supersymmetry and dark matter, and measuring the properties of quantum chromodynamics (QCD).  
Because this task involves highly-correlated and high-dimensionality inputs, it is an active area of R\&D for ML algorithms in particle physics.  
Various representations of the data have been considered, including fixed 2D images, variable length sets, and graphs.

In this case study, we consider the task of classifying collimated decays of top quarks in a jet from more common jets originating from lighter quarks or gluons.
There are many ML approaches to this challenge in the literature~\cite{Kasieczka:2019dbj} and a public dataset, developed from one of these studies, has been created for comparison~\cite{Butter:2017cot,topdataset}.
The \textsc{Pythia8}~\cite{Sjostrand:2014zea,Skands:2014pea} generator is used to produce fully hadronic $\ttbar$ events for signal (known as``top quark jets'') and QCD dijet events for background (known as ``QCD jets'') produced in 14 TeV proton-proton collisions. 
No multiple parton interactions or pileup interactions are included and their inclusion would require improving our neural network model. 
\delphes~\cite{deFavereau:2013fsa} with the ATLAS detector configuration is used to simulate detector effects.
The \delphes~E-flow candidates are clustered using \textsc{FastJet}~\cite{fastjet:1,fastjet:2} into anti-$\kt$~\cite{Cacciari:2008gp} jets with size parameter $R=0.8$.
Jets with transverse momentum ($\pt$) between 550 and 650~GeV and $\abseta < 2$ are selected where $\eta$ is the pseudorapidity.
Top quark jets are required to satisfy generator-level matching criteria: the jet must be matched to a parton-level top quark and all of its decay products within $\Delta R = 0.8$, where $\Delta R = \sqrt{(\Delta\eta)^2 + (\Delta\phi)^2}$ and $\phi$ is the azimuthal angle. 
Up to 200 jet constituent four-momenta are stored.

\begin{figure*}[tbh!]
\centering
\includegraphics[width=0.44\linewidth,clip=true, viewport = 0 80 864 756]{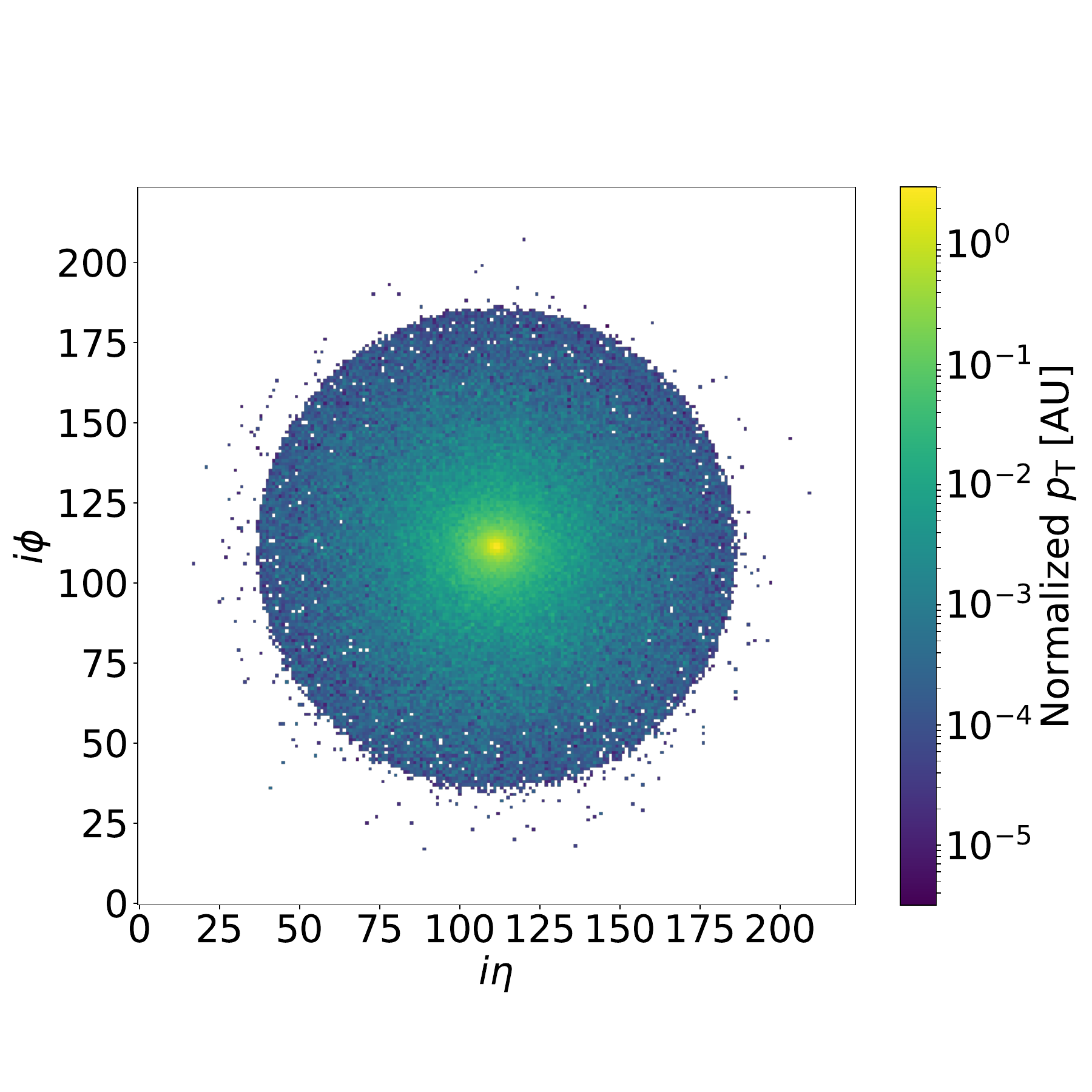}
\includegraphics[width=0.44\linewidth,clip=true, viewport = 0 80 864 756]{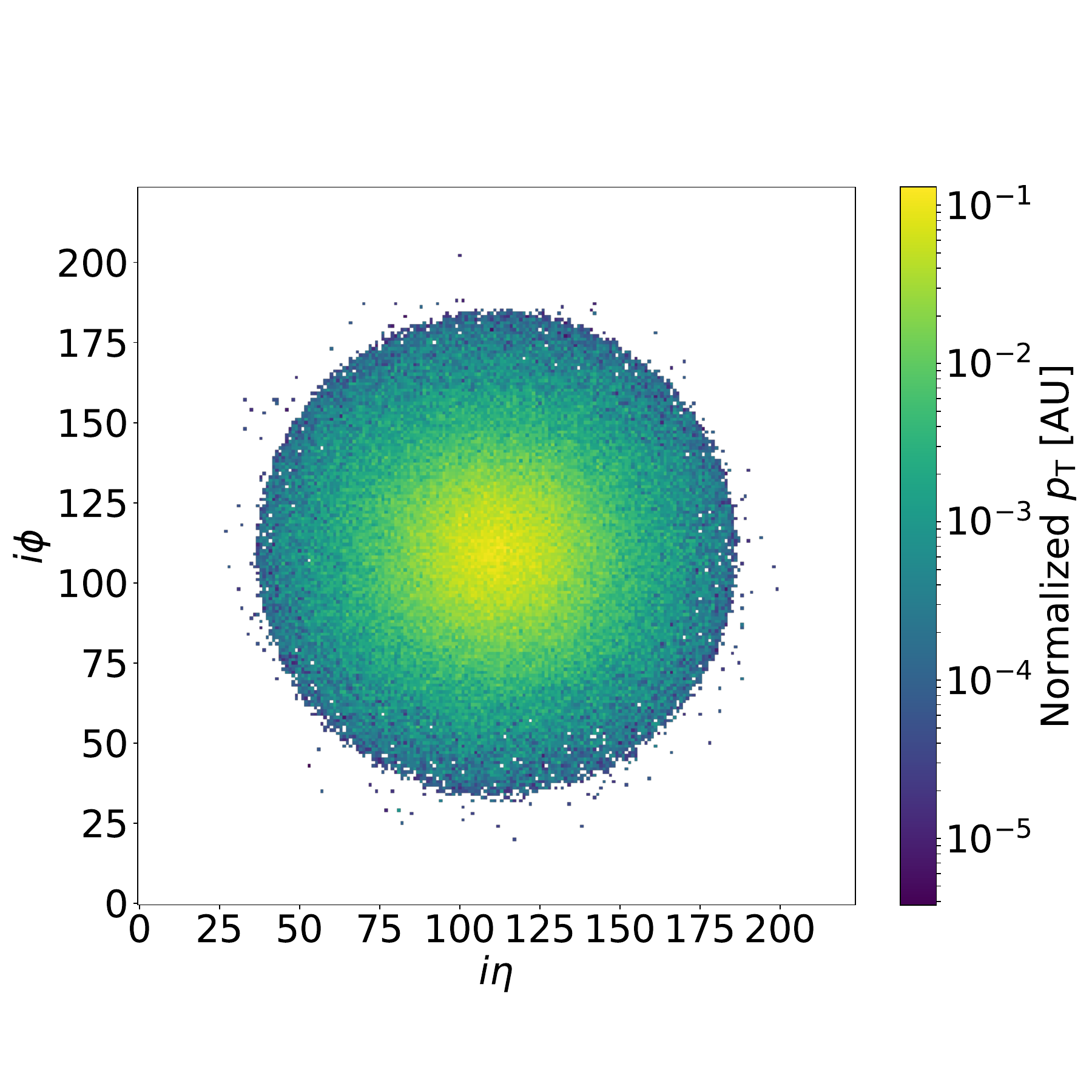}
\caption{A comparison of QCD (left) and top (right) jet images averaged over 5,000 jets.}
\label{fig:substr-images}
\end{figure*}

The Brainwave platform allows the use of custom weights for specific applications computed by training predefined CNNs.
In this training, we treat the jets as 2D grayscale images in the $\eta$-$\phi$ plane and send them as input to the \resnet algorithm.
Jet images are created by summing jet constitutent $\pt$ in a 2D grid of $224\times 224$ in $\eta$ and $\phi$ units from $-1.2$ to $1.2$ centered on the jet axis~\cite{Huilin}.
In order to apply the standard \resnet architecture, the images are normalized such that each image has a range between $0$ and $225$
and duplicated 3 times, once for each RGB channel.
We illustrate the images for QCD and top quark jets in Fig.~\ref{fig:substr-images} where the images are averaged over 5,000 jets.  
Top quark jets have a 3-prong nature which manifests as a broader radiation pattern when averaged over many jets.  

\begin{sloppypar}For our specific task,  after the primary \resnet featurizer we add our own custom classifier, which comprises one fully connected layer of width $1024$ with ReLU~\cite{ReLU} activation and another fully connected layer of width $2$ with softmax activation.
The training dataset contains about 1.2 million events while the validation and test datasets each have approximately 400,000 events.  
The training is performed by minimizing the categorical cross-entropy loss function using the Adam algorithm~\cite{DBLP:journals/corr/KingmaB14} with an initial learning rate of $10^{-3}$ and a minibatch size of $64$ over $10$ epochs on an NVIDIA Tesla V100 GPU.
The best model is chosen based on the smallest average loss evaluated on the validation dataset.
The training for this particular \resnet model is unique because there is a particular {\it quantized} version of \resnet that needs to be ``fine-tuned,'' or trained with a smaller learning rate.
The quantized model is initialized using the weights from the trained floating point model and trained with an initial learning rate of $10^{-4}$ and a minibatch size of $32$ for 10 additional epochs.
Finally, as the quantized model evaluated with the Brainwave FPGA service differs numerically from the quantized model evaluated on the local GPU, an additional fine-tuning is applied to the classifier after evaluating the \resnet features on Brainwave. 
This fine-tuning of the classifier layers is performed over $100$ epochs using the validation data with the Adam algorithm, an initial learning $10^{-4}$, and a batch size of $128$.
On a single V100 GPU, the initial floating point training time is approximately 1.5 hours per epoch while the ``fine-tuned'' training is approximately 4 hours per epoch.  The classifier layer training is significantly faster, only minutes per epoch.  
\end{sloppypar}

\begin{figure}[tbh!]
\begin{center}
\includegraphics[width=0.98\linewidth]{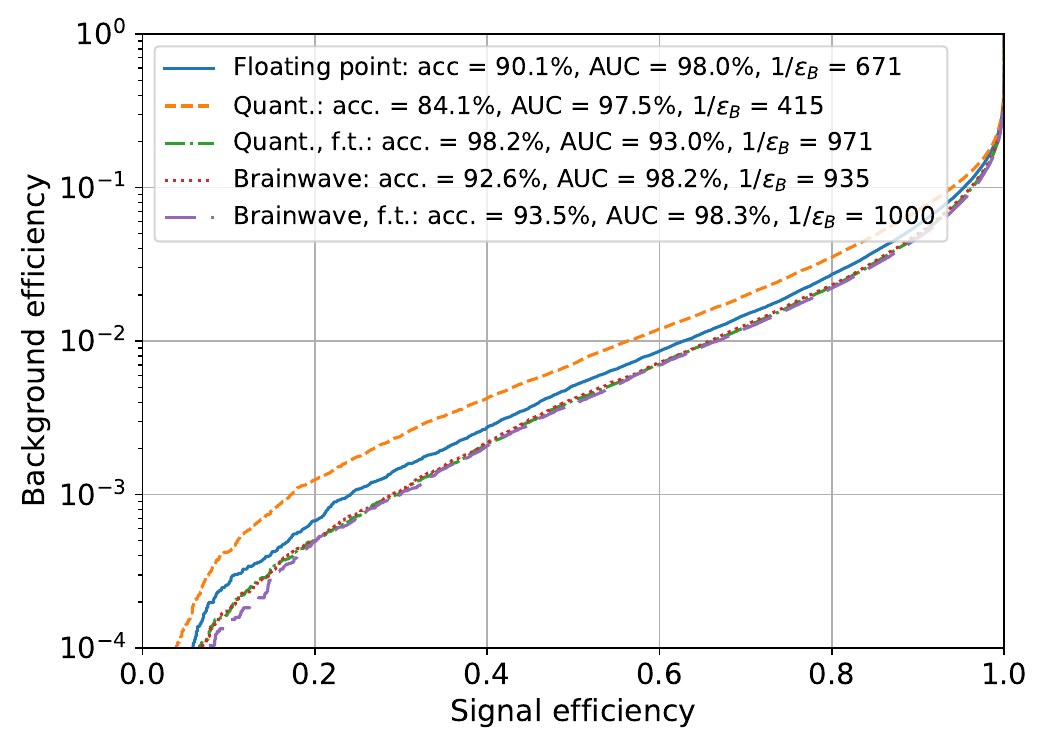}
\end{center}
\caption{The ROC curves showing the performance of the floating point and quantized versions (before fine-tuning, after fine-tuning, and using the Brainwave service) of the \resnet top tagging model.}
\label{fig:substr-ROC}
\end{figure}

After training, we evaluate the performance of our trained \resnet top tagger.  
The receiver operator characteristic (ROC) curve is a graph of the false positive rate (background QCD jet efficiency) as a function of the true positive rate (top quark jet efficiency.)
It is customary to report three metrics for the performance of the network on the top tagging dataset: model accuracy, area under the ROC curve (AUC), and background rejection power at a fixed signal efficiency of 30\%, $\bkgrej$.  
Fig.~\ref{fig:substr-ROC} shows the ROC curve comparison for the transfer learning version of \resnet as well as the fully retrained featurizer with custom weights.
In Table~\ref{tab:perf}, the accuracy, AUC, and $\bkgrej$ values are listed for each model considered.  
The performance of the retrained \resnet compared to other models developed for this dataset is state-of-the-art; the best performance is ${\bkgrej \approx 1000}$.

\begin{table}
\centering
\begin{tabular}{c|cccc}
Model & Accuracy & AUC & $\bkgrej$ \\
\hline
Floating point & 0.9009 & 0.9797 & 670.8 \\
Quant. & 0.8413 & 0.9754 & 414.6 \\
Quant., f.t. & 0.9296 & 0.9825 & 970.7\\
Brainwave & 0.9257 & 0.9821 & 934.8\\
Brainwave, f.t. & 0.9348 & 0.9830 & 999.6
\end{tabular}
\caption{\label{tab:perf} The performance of the evaluated models on the top tagging dataset.}
\end{table}

One other consideration in this study is the size of the model.
The typical particle physics models used for top tagging are often several orders of magnitude smaller than \resnet in terms of the numbers of parameters and operations.
However, it should be noted that the best-performing models to date ({\tt ResNeXt50} and a directed graph CNN)~\cite{Huilin,Kasieczka:2019dbj} are within a factor of a few in size with respect to the \resnet model.  
We emphasize here that this study is a proof-of-concept for the physics performance and that there are many other very challenging, computationally intensive algorithms where machine learning is being explored.  
We anticipate that for these looming challenges, the size of the models will continue to grow to meet the demands of new experiments.

\subsection{Neutrino flavor identification at \texorpdfstring{\nova}{NOvA}}
\label{sec:nova}
Neutrino event classification can also benefit from accelerating the inference of large ML models.
In this section, due to a lack of publicly available neutrino datasets, we do not
fully quantify the performance of a particular model. 
Instead, we present a workflow to demonstrate that this work is applicable beyond the LHC.

We illustrate the type of classification task needed for neutrino
experiments by using simulated neutrino events and cosmic data from
the {\nova} experiment. {\nova}
pioneered the application of convolutional neural networks (CNN) in particle physics in 2016 by
becoming the first experiment to use a CNN in a published result~\cite{Aurisano:2016jvx,Adamson:2017gxd}.
In our study, we use \emph{transfer learning} with \resnet to distinguish
between the different detector signatures associated with various
neutrino interaction types and associated backgrounds.
We extract features from neutrino interaction events using the \resnet featurizer (pre-trained using the ImageNet dataset~\cite{imagenet_cvpr09}) and retrain the final fully connected classifier layers to perform neutrino event classification. 
Specifically, 500,000 simulated neutrino events with cosmic data overlays
were used for training, with the following five categories:
charged current electron neutrino, charged current muon neutrino,
charged current tau neutrino, neutral current neutrino interactions,
and cosmic ray tracks. These events are highly amenable to
classification by CNN architectures such
as \resnet.

\begin{figure*}[tbh!]
\centering
\includegraphics[width=0.26\linewidth]{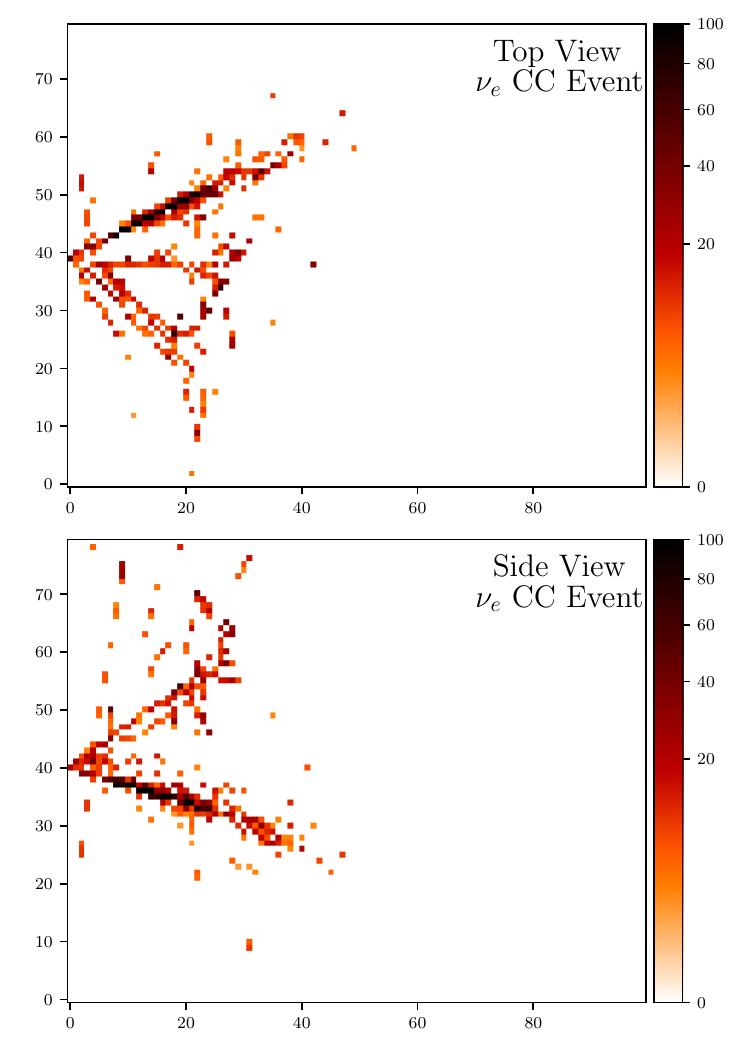}
\includegraphics[width=0.26\linewidth]{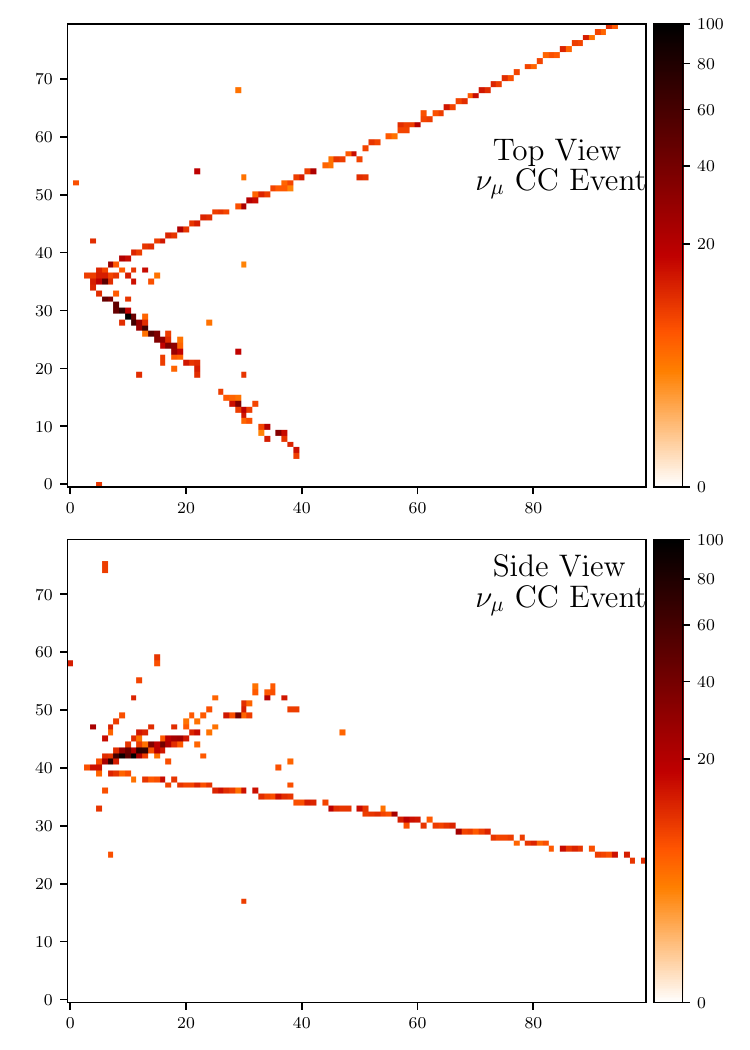}
\includegraphics[width=0.26\linewidth]{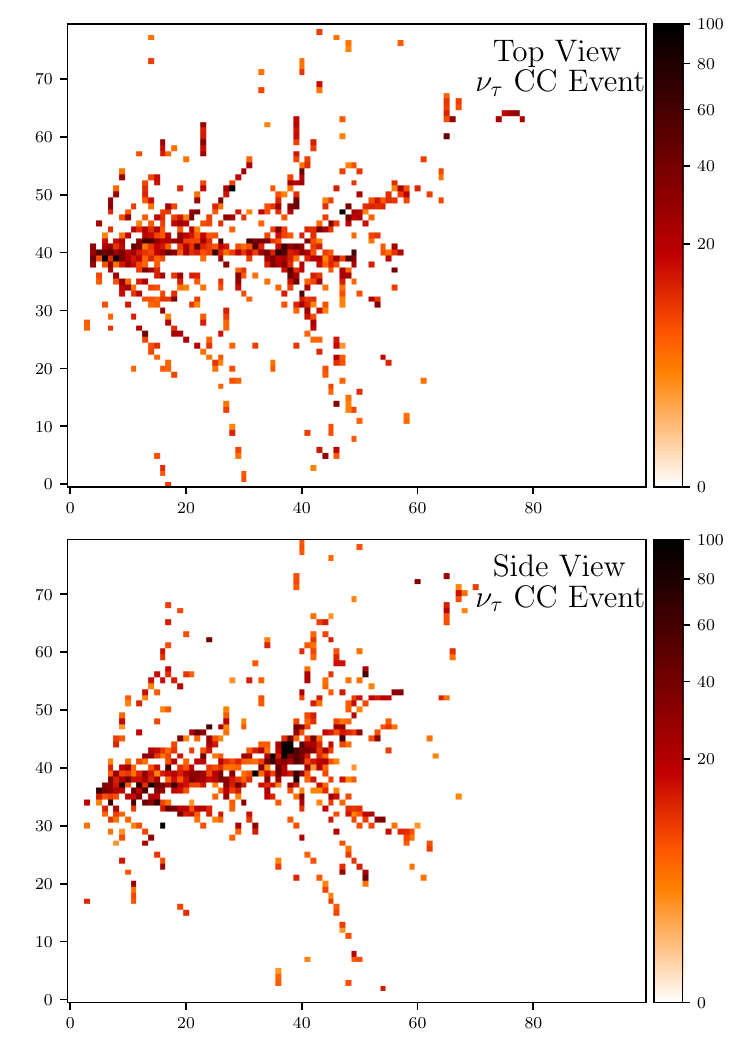}
\caption{
\label{fig:nova} Example visualizations of simulated neutrino events
correctly classified by our \resnet model with probability greater than 0.9:
electron neutrino (left),
muon neutrino (middle), and tau neutrino (right). The top and bottom
rows are the top and side views from the {\nova} detector.
(NOvA's beam energy and
baseline prohibit long baseline tau neutrino appearance searches, but
the event is shown for illustration purposes.)}
\end{figure*}

We then applied the transfer learning \resnet model to a separate test
set of 150,000 events. As a visual example, we show three
simulated neutrino
interaction type events in Fig.~\ref{fig:nova} that are selected with
probability, larger than 0.9. On the left (middle, right) is an example
event originating from an electron (muon, tau) neutrino charged
current interaction. While the optimal use of ML to
improve neutrino event reconstruction and classification is an active
area of research, 
the most successful approach thus far employs CNN architectures, which work well with the homogeneous nature of the neutrino
detectors. 
While the transfer learning approach does not yield state-of-the-art performance for neutrino event classification, we expect that a full retraining of \resnet would be more successful, which is the subject of future work.

Current neutrino experiments, including {\nova} and others, are
potentially exciting applications of coprocessors as a service. A large fraction of
their event reconstruction time is already consumed by inference of
large CNNs~\cite{himmel}. Therefore, they stand to
gain significantly from accelerating network inference. The
approach outlined in Section~\ref{sec:sonic} could provide a non-disruptive solution to accelerate
neutrino computing performance in the present as well as in the future.

\section{Heterogeneous computing as a service}
\label{sec:sonic}
\subsection{FPGA coprocessors as a service}

\begin{figure*}[t]
\centering
\includegraphics[width=0.8\textwidth]{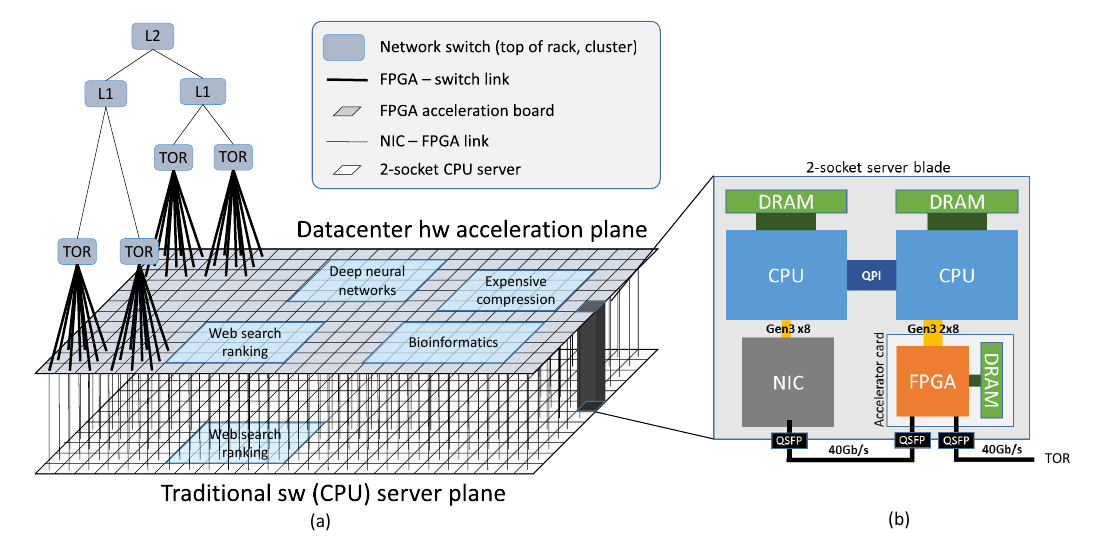}
\caption{A schematic of the Microsoft Brainwave acceleration platform~\cite{configurable-cloud-acceleration}.}
\label{fig:bw}
\end{figure*}

In this study, we explore how to integrate heterogeneous computing solutions into the particle physics computing paradigm.
The jet physics model developed in the previous section is used as a specific motivating example.
In our work, we benchmark the recently released Microsoft Brainwave platform which performs acceleration with Intel Altera FPGAs~\cite{configurable-cloud-acceleration}. 
FPGAs as a computing solution offers a combination of low power usage, parallelization, and programmable hardware.
Another important aspect of FPGA inference for the particle physics community, compared to GPU acceleration, is that batching is not required for high performance; FPGA performance is not diminished for serial processing.
The Brainwave system, in particular, has demonstrated the use of FPGAs in a cloud system to accelerate ML inference at large scale~\cite{configurable-cloud-acceleration}.
In Fig.~\ref{fig:bw}, we show a schematic of the Brainwave system from Ref.~\cite{configurable-cloud-acceleration}, which illustrates its cloud-scale configurable FPGA setup for acceleration.
The Brainwave system includes interconnectivity of the FPGA acceleration elements and a direct connection to the network, which runs in parallel to the CPU-based software plane.  
The performance of other available acceleration hardware systems will be explored in future work.


Deploying ML algorithms in particle physics have two particularly interesting benefits to the computing model:
\begin{itemize}
	\item By considering ML algorithms, we can greatly benefit from developments outside of the field of particle physics.  Industry and academic investment in ML is growing rapidly, and there is a vast amount of research on specialized hardware for ML that could be utilized within the community.
	\item Often, ML algorithms are quite parallelizable, making them amenable to acceleration on specialized hardware. For some physics-based algorithms, this is not possible, while for others it could require substantial investment to rewrite for new, often changing computing hardware.
\end{itemize}
We, therefore, focus on ML acceleration in our study. To capitalize on the ML-focused hardware developments, we rely on the continued research and development of ML applications for particle physics tasks.
This is an active area of research with growing interest, as indicated by recent work across many neutrino and collider experiments~\cite{Radovic:2018dip,Albertsson:2018maf} and initiatives such as the HEP.TrkX project~\cite{heptrkx} and the Tracking ML Kaggle Challenge~\cite{trackml}.
Additionally, ML has the potential to provide event simulation~\cite{Paganini:2017dwg}, another computationally intensive part of the chain.

\begin{figure}
\begin{center}
\includegraphics[width=0.9\linewidth]{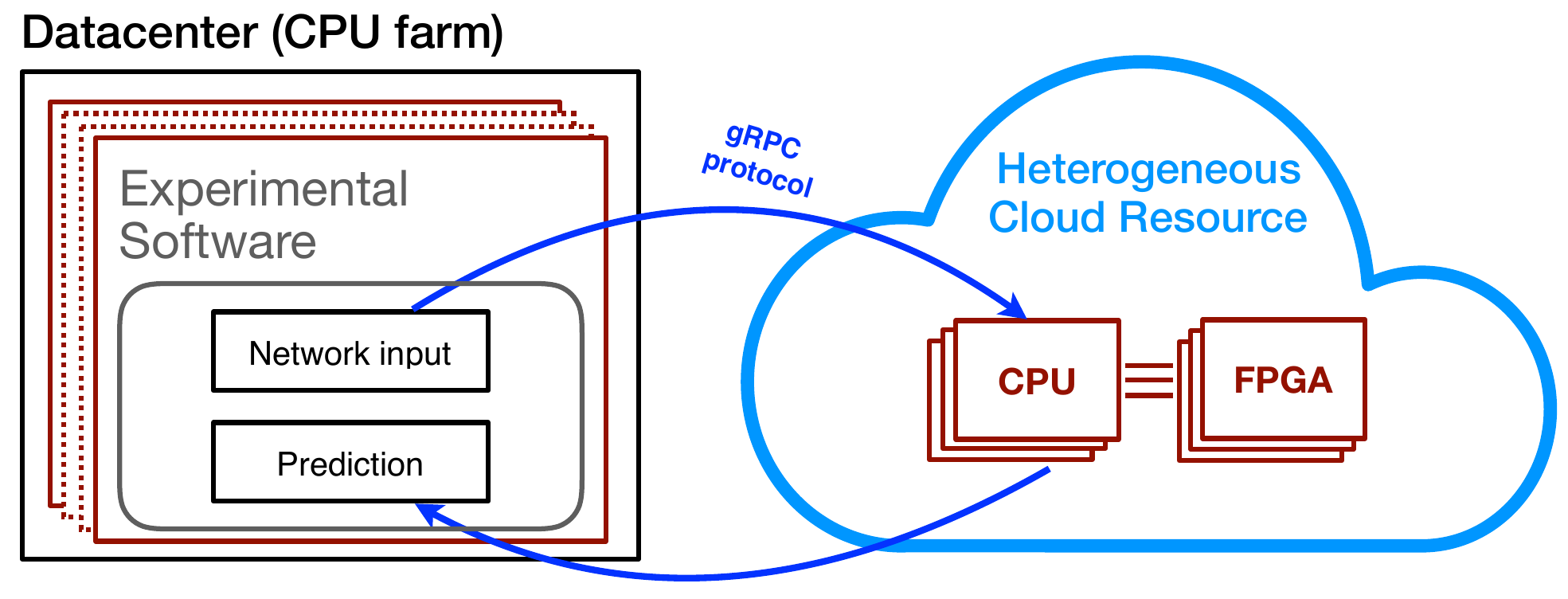}
\end{center}
\caption{An illustration of FPGA-accelerated ML cloud resources integrated into the experimental physics computing model as a service.}
\label{fig:soniccloud}
\end{figure}

One challenge is to integrate FPGA coprocessors into the computing model without disrupting the current multithreaded paradigm, where several modules process an event in parallel.
A natural method for integrating heterogeneous resources is via a network service.
This client-server model is flexible enough to be used locally by a single user or within a computing farm where a single thread communicates with the server.
In the particular case investigated here, we use the \gRPC package~\cite{gRPC}, an open-source Remote Procedure Call (RPC) system developed initially by Google, interfaces with the Brainwave system.
\gRPC uses protocol buffers (\protobuf)~\cite{protobuf} for data serialization and transmission.
This setup defines a communication method between the FPGA coprocessor resources and an experiment's primary computing CPU-based datacenters.
This is illustrated in Fig.~\ref{fig:soniccloud} where a module running on a CPU farm performs fast inference of a particular ML algorithm via \gRPC.
First, we test the performance of a single task which makes a request to a single {\it cloud} service which performs a {\it remote}\footnote{we refer synonymously to a {\it cloud} service being accessed {\it remotely}} access to the Brainwave platform.  
However, scaling up the number of requests is natural for the Brainwave system, which is capable of load balancing of service requests.

\begin{figure}[tbh!]
\begin{center}
\includegraphics[width=0.7\linewidth]{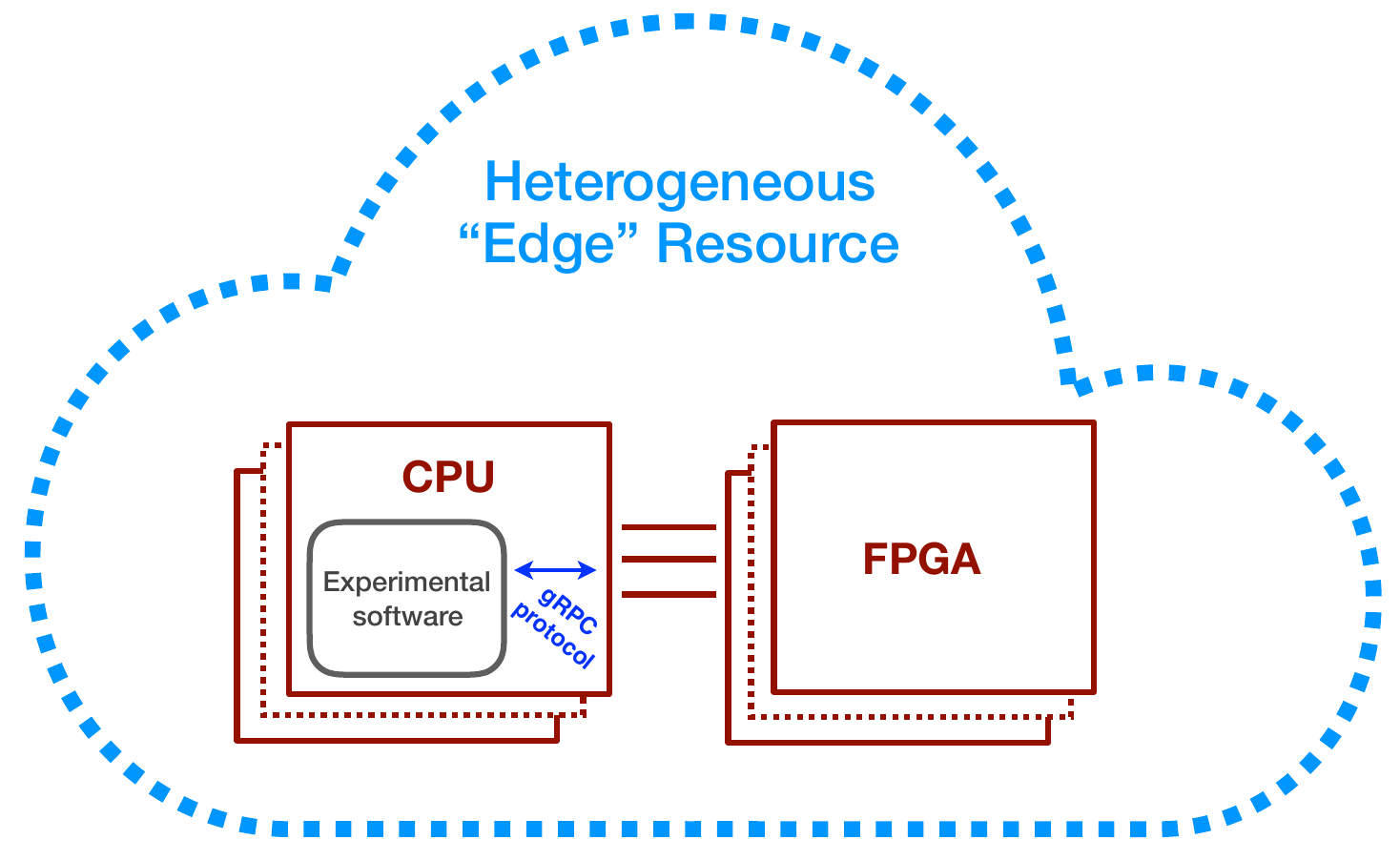}
\end{center}
\caption{An illustration of FPGA-accelerated ML edge resources integrated into the experimental physics computing model as a service.}
\label{fig:sonicedge}
\end{figure}

One may also consider a case where the FPGA coprocessor resources are located at the same datacenter, {\it on-premises}, as the CPUs, as a so-called {\it edge} resource\footnote{we refer synonymously to a {\it edge} service being accessed {\it on-premises, or on-prem}}.
This is illustrated in Fig.~\ref{fig:sonicedge}.
In this scenario, the same \gRPC interface protocols are used to communicate with the FPGA hardware, and the software access for fast inference is unchanged.  
To benchmark this scenario, we run our application on a virtual machine (VM) in the cloud datacenter.
Results comparing both these scenarios with other hardware from the literature are presented in Section~\ref{sec:results}.

\subsection{Particle physics computing model with services}

For our demonstration study, we use the CMS experiment software framework, \cmssw~\cite{CMSSW}. 
This software uses Intel Thread Building Blocks~\cite{IntelTBB} for task-based multithreading.
A typical module, such as those depicted in Fig.~\ref{fig:edm}, has a \produce function that obtains data from an event, operates on it, and then outputs derived data.
This pattern assumes that all of the operations occur on the same machine.

\begin{figure}[tbh!]
\begin{center}
\includegraphics[width=0.9\linewidth]{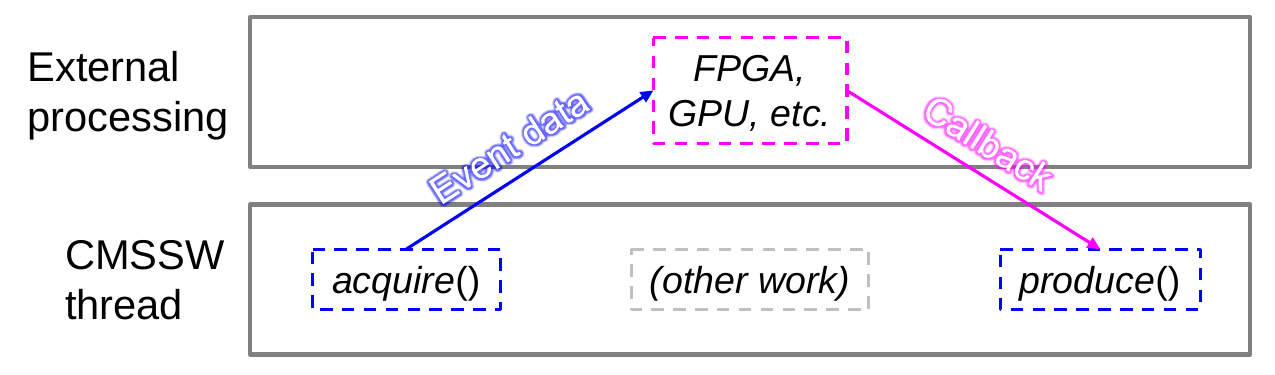}
\end{center}
\caption{A diagram of the \externalwork feature in \cmssw, showing the communication between the software and external processors such as FPGAs.}
\label{fig:ew}
\end{figure}

Our goal is to utilize the Brainwave hardware as a service to perform inference of a large ML model such as \resnet. 
Within \cmssw, a hook to the \gRPC system is established using a special feature called \externalwork.
Optimal use of both CPU and heterogeneous computing resources requires that requests be transmitted asynchronously, freeing up a CPU thread to do other work rather than forcing it to wait until a request is complete.
The \externalwork pattern accomplishes this by splitting the simpler pattern described above into two steps.
The first step, the \acquire function, obtains data from an event, launches an asynchronous call to a heterogeneous resource, and then returns.
Once the call is complete, a callback function is executed to place the corresponding \produce function for the module back into the task queue.
This is depicted in Fig.~\ref{fig:ew}.

In this case, the event data provided to the service is a \tensorflow tensor with the appropriate size ($224 \times 224 \times 3$) for inference with \resnet.
A list of the classification results is returned back to the module, which employs \externalwork.  
For simplicity, we refer to the full chain of inference as a service within our experimental software stack as ``Services for Optimized Network Inference on Coprocessors'' or \sonic~\cite{SonicSW}.

\section{Computing performance and results}
\label{sec:results}
\subsection{Brainwave performance} 

We benchmark the performance of the \sonic package within \cmssw, measuring the total end-to-end latency of an inference request using Brainwave. 
In a simple test, we create an image from a jet (as described in Sec.~\ref{sec:physics}) from a simulated CMS dataset.
We take reconstructed particle candidates and combine them as pixels in a 2D grayscale image tensor input to the \resnet model (as in Sec.~\ref{sec:results-trainLHC}).

\begin{sloppypar}We perform two latency tests: \remote and on-premises or \onprem.
The \remote test communicates with the Brainwave system as a cloud service, as illustrated in Fig.~\ref{fig:soniccloud}.  
For this test, we execute our experimental software, \cmssw, on the local Fermilab CPU cluster (Intel Xeon 2.6 GHz) in Illinois, US, and communicate via \gRPC with the service located at the Azure East 2 Datacenter in Virginia, US. 
The \onprem tests are executed at the same datacenter as the Brainwave FPGA coprocessors.
We run a VM in the Azure East 2 Datacenter, deploying \cmssw inside a Docker container, and communicate with the FPGA coprocessors located in the same facility.
\end{sloppypar}

\begin{figure}[tbh!]
\begin{center}
\includegraphics[width=0.8\linewidth]{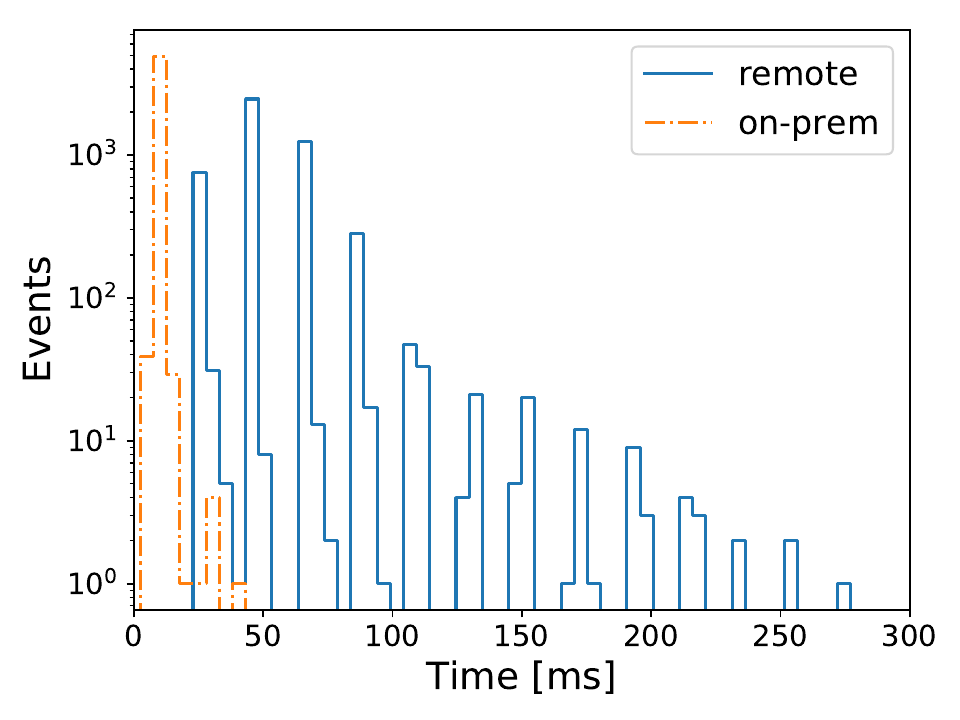}\\
\includegraphics[width=0.8\linewidth]{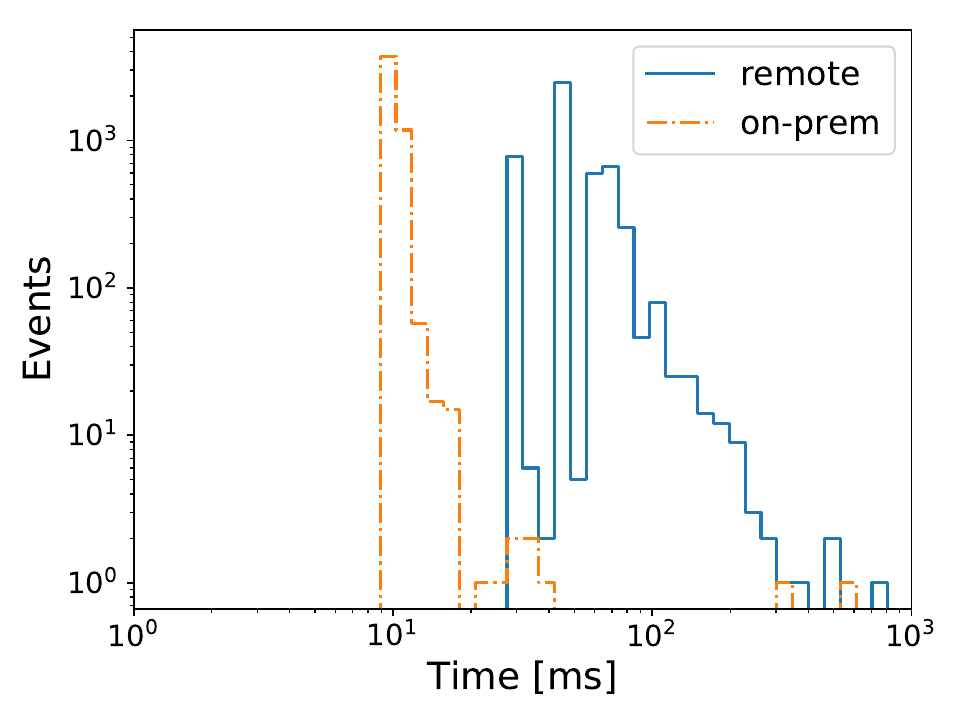}
\end{center}
\caption{Total round trip inference latencies for \resnet on the Brainwave system both \remote and \onprem.  The top plot is linear in time and the bottom plot is logarithmic in time. }
\label{fig:bwres}
\end{figure}

We measure the total round-trip latency of the inference request as seen by \cmssw, starting from the transmission of the image and ending with the receipt of the classification results.  
The latencies are shown in Fig.~\ref{fig:bwres} for a linear latency scale (top) and a logarithmic latency scale (bottom).
The \onprem performance is shown in orange, with a mean inference time of 10~ms, and the \remote performance is shown in blue, with a mean inference time of 60~ms.
From internal Brainwave timing tests, the featurizer inference step performed on the FPGA takes 1.8~ms and the classifier inference step performed on the CPU is similar. The remaining time in the 10~ms is primarily used for network transmission.

The \remote performance can be as fast as 30~ms with a median value of 50~ms, and there are long tails out to hundreds of ms at the per-mille level.
The measured latency is strongly dependent on network conditions which can cause the structures seen in Fig.~\ref{fig:bwres}.
Due to the speed of light, there is a hard physical limit in the transmission time of the signal to the Azure East 2 Datacenter and back to Fermilab, which we estimate to be around 10~ms.  
The physical distance between the experimental computing cluster and the remote datacenter will limit any cloud-based inference speeds.

\begin{figure}[tbh!]
\begin{center}
\includegraphics[width=0.8\linewidth]{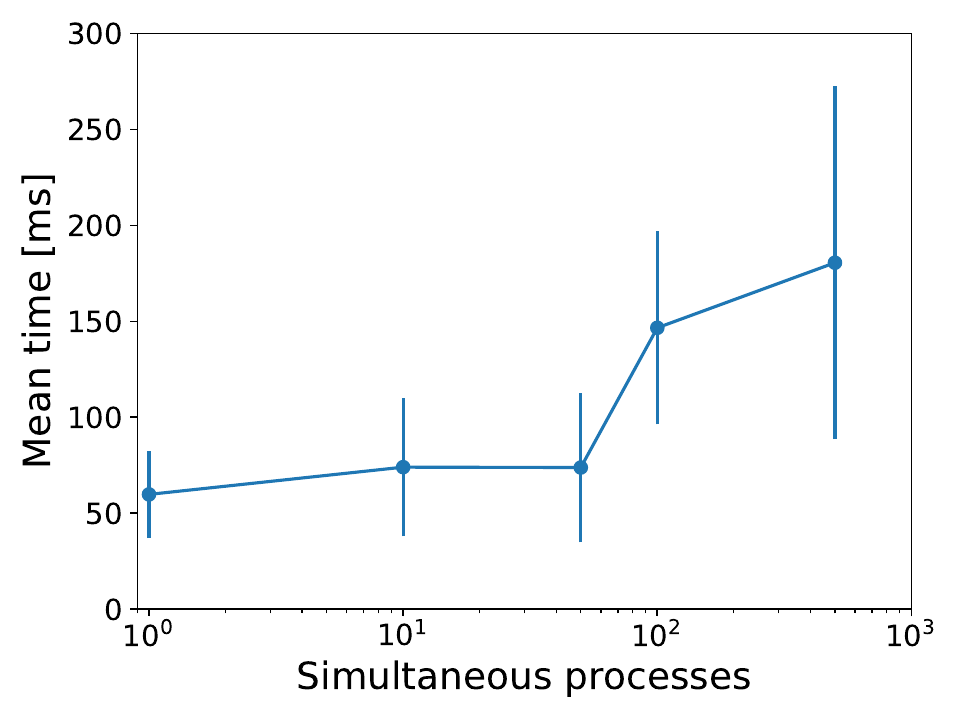}\\
\includegraphics[width=0.8\linewidth]{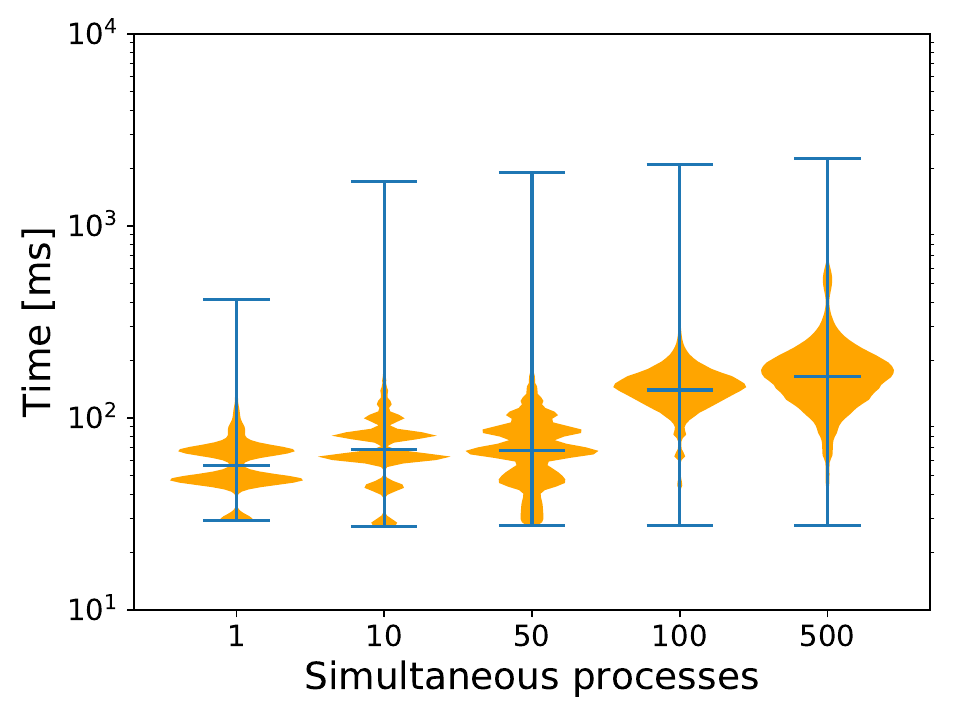}
\end{center}
\caption{Top: Mean round trip inference latencies for \resnet on the Brainwave system for different numbers of simultaneous processes. The error bars represent the standard deviation.
Bottom: The full distributions displayed in ``violin'' style. The vertical bars indicate the extrema. The horizontal axis scale is arbitrary.}
\label{fig:stress}
\end{figure}

After comparing the \remote versus \onprem latency, we performed a scaling test to estimate how many coprocessor services would be needed to support large-scale deployment in a production environment.
A given number of simultaneous processes were run using the batch system at Fermilab and the round-trip latency was measured. All jobs connected to a single Brainwave service.
This test corresponds to a ``worst-case'' estimation of the scaling of a single service because each process only executed the Brainwave test module that performs inference on jet images.
In an actual production process, the test module would run alongside many other modules (see Fig.~\ref{fig:edm}), greatly reducing the probability of simultaneous requests to the cloud service.
The results of the test are shown in Fig.~\ref{fig:stress}. The mean, standard deviation, and long tail for the round trip latency all tend to increase with more simultaneous jobs, but only moderately.
It should also be noted that some calls timed out during the largest-scale test with 500 simultaneous processes, leading to a failure rate of 1.8\%, while the other tests had zero or negligible failures.

\begin{figure}[tbh!]
\begin{center}
\includegraphics[width=0.8\linewidth]{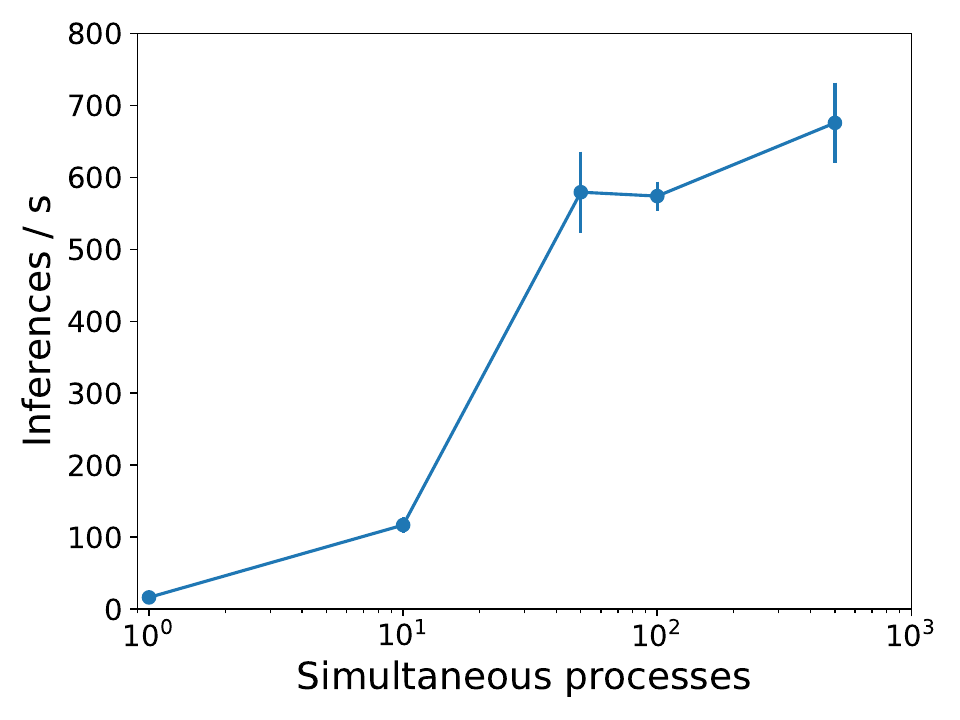}\\
\includegraphics[width=0.8\linewidth]{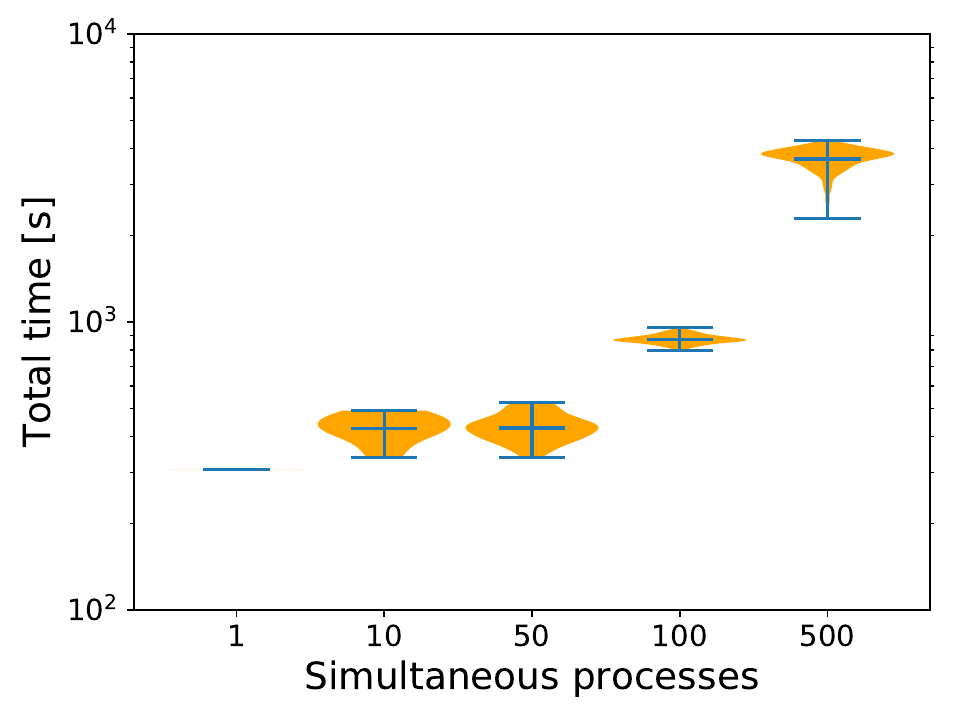}
\end{center}
\caption{Top: Throughput of the FPGA service as the number of inferences per second for different numbers of simultaneous processes. The error bars represent the standard deviation.
Bottom: mean total time and distribution (in seconds) to process 5000 jet images through \resnet on the Brainwave system for different numbers of simultaneous processes. The vertical bars indicate the extrema. The horizontal axis scale is arbitrary.}
\label{fig:stress-fulltime}
\end{figure}

We also measure the throughput based on the total time for each simultaneous process to complete serial processing of 5000 jet images.
These results are shown in Fig.~\ref{fig:stress-fulltime}.
Though the round trip latency for a single request has a large variance, the total time to process the full series of images is remarkably consistent.
This demonstrates the efficient load balancing performed by the Brainwave server. 

\begin{sloppypar}With the total time measured for all simultaneous processes to complete, we can compute the total throughput of the Brainwave service.  
Recall from above that while the cloud service inference round trip latency is 60~ms, on average, the latency for the featurizer inference on the FPGA itself is approximately 1.8~ms.
When we run multiple simultaneous CPU processes that all send requests to one service, we fully populate the pipeline of data streaming into the service.
This keeps the FPGA occupied, increasing its duty cycle and the total inference throughput of the service.
This is illustrated in Fig.~\ref{fig:stress-fulltime}, where we show the throughput of the service in inferences per second as a function of the number of simultaneous CPU processes accessing the service.
As the number of simultaneous processes increases, the number of inferences per second increases, because of the increased pressure on the pipeline of the FPGA service.  
The mean latency, shown in Fig.~\ref{fig:stress}, does not degrade much as the number of simultaneous jobs increases from 1 to 50, while the throughput increases by a factor of nearly 40 (600 inferences per second).
The throughput of the service plateaus at around 650 inferences per second; it is limited by the inference time on the FPGA that is, at best, 1.8~ms.
From these studies, we find that it is more efficient and also more cost-effective to have multiple simultaneous CPU processes connect to a single FPGA service.
\end{sloppypar}

The ratio of simultaneous processes to FPGA services is dependent on the other tasks in the process; typical physics processes run many modules.
The tests we have performed are the most pessimistic scenario because each process only executes the Brainwave test module.
Therefore, in more realistic workloads where many tasks are run per process and a majority of those tasks run on the CPU, we expect that one FPGA service will be able to serve one model for many more than 50 simultaneous CPU processes.  Detailed studies of these more realistic workloads will be performed in the future.

\subsection{CPU/GPU comparisons} 

Next, we compare the performance of the Brainwave platform to CPU and GPU performance for the same \resnet model.
Such comparisons can be greatly affected by many details of the entire computing stack and vary widely even within the literature.
Nonetheless, to get a sense of the relative performance, we perform two types of tests.  
First, we do our own standalone python benchmark tests with the azure-ML implementation of \resnet as well as the \tensorflow implementation of the \resnet model. Here, we verify our results against the literature.
While many more detailed studies exist, these benchmarks validate our numbers against other similar tests.  
Second, we import the \resnet model file provided by Brainwave into \cmssw and perform inference on the local CPU with the version of \tensorflow currently in the \cmssw release~\footnote{It takes significant effort to adapt \tensorflow to be compatible with the multithreading pattern used in \cmssw, and hence the latest version of \tensorflow is usually not available to be used in the experiment's software.}.

\begin{figure}[tbh!]
\begin{center}
\includegraphics[width=0.9\linewidth]{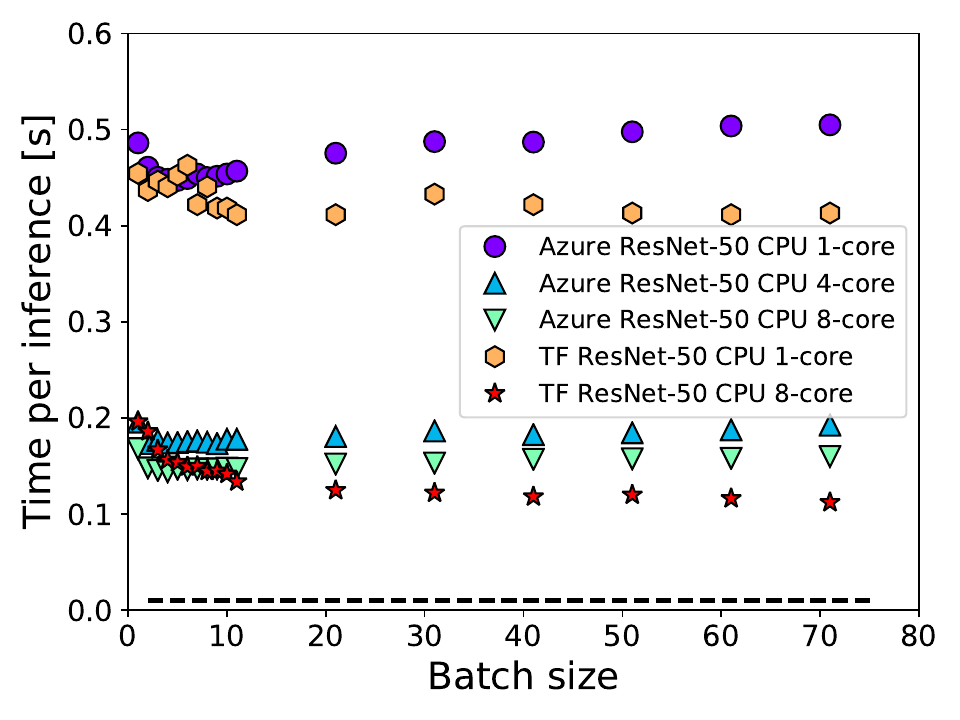} \\
\includegraphics[width=0.9\linewidth]{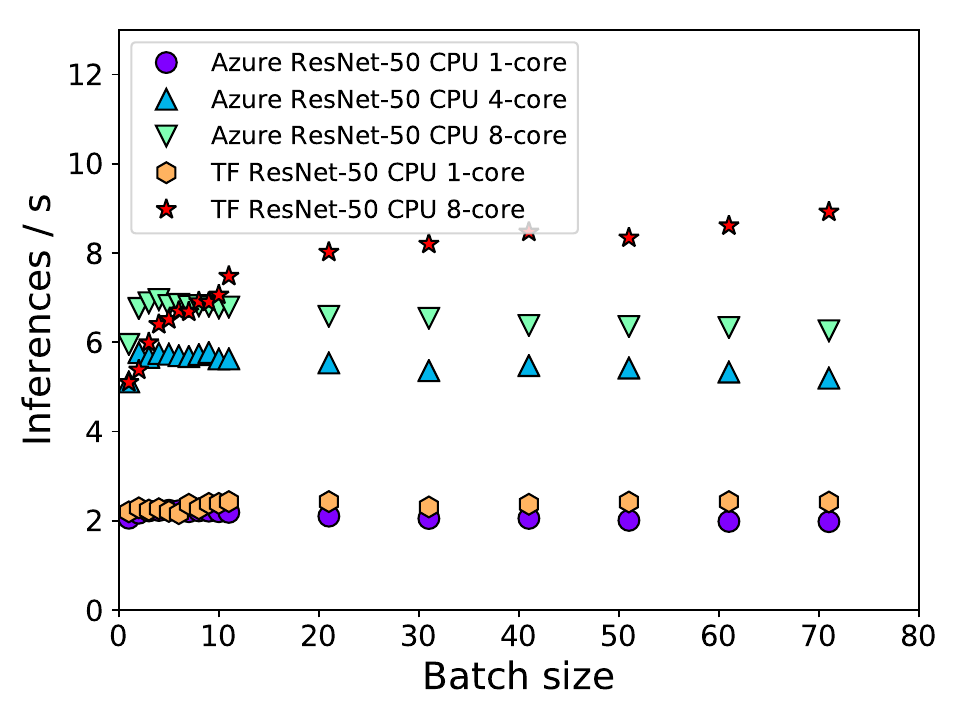}
\end{center}
\caption{Standalone CPU inference time per image (top) and images per second (bottom) as a function of batch size for the \tensorflow official \resnet model compared with the Azure \resnet model. The dashed line indicates a time of 10~ms, consistent with the \onprem inference time of the Brainwave system. }
\label{fig:cpu}
\end{figure}

The standalone python benchmark results for CPUs are presented in Fig.~\ref{fig:cpu}.
The CPU used in these tests is an Intel i7 3.6~GHz.
For the CPU, we compare the number of cores used for either the Brainwave implementation of \resnet or the conventional {\tensorflow} \resnet.
The performance is shown versus the image batch size; particle physics applications can vary in their batch sizes typically from 1 to 100.
As expected, the performance is stable versus batch size. For both models, we observe roughly the same inference time, ranging from roughly 180~ms to 500~ms.
Additionally, we observe that the model inference time is close to optimal when using 4 cores, with small improvements beyond.

\begin{figure}[tbh!]
\begin{center}
\includegraphics[width=0.9\linewidth]{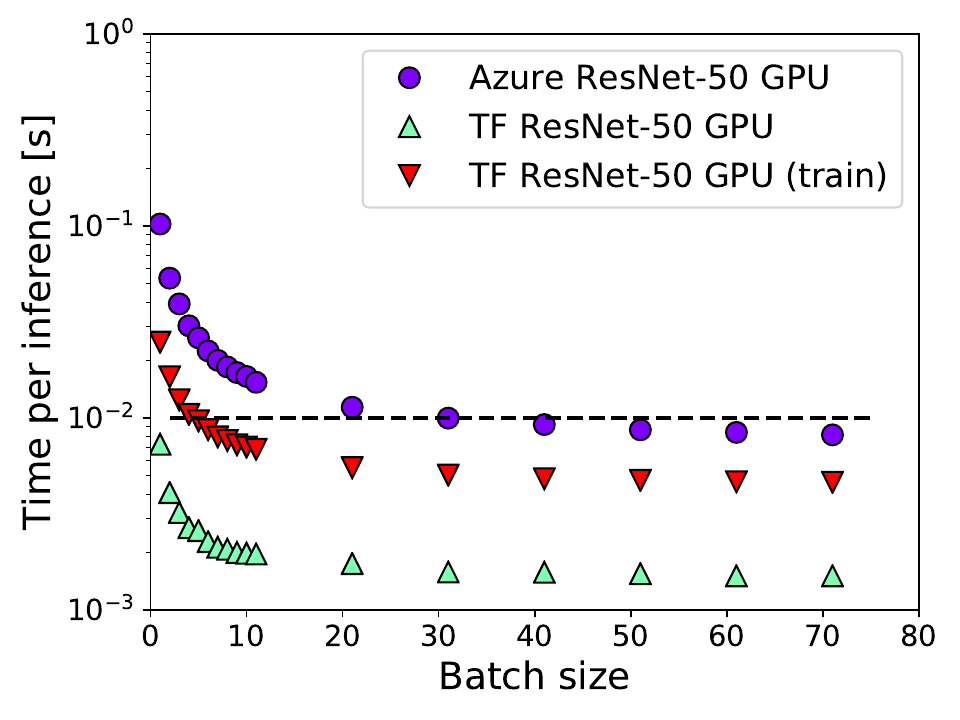} \\
\includegraphics[width=0.9\linewidth]{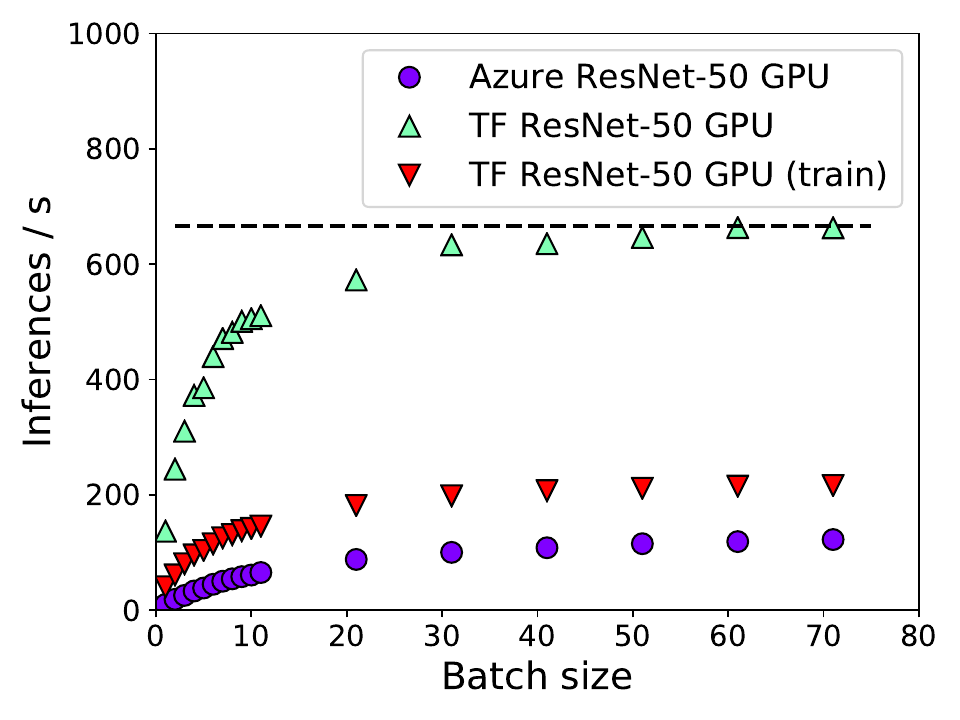}
\end{center}
\caption{Standalone GPU inference time per image (top) and images per second (bottom) as a function of batch size for the \tensorflow official \resnet model compared with the Azure \resnet model. The dashed line indicates a time of 10~ms, consistent with the \onprem inference time of the Brainwave system. }
\label{fig:gpu}
\end{figure}

\begin{sloppypar}Figure~\ref{fig:gpu} shows the inference times on GPUs.
It is important to note that the GPU used in these tests, an NVidia GTX 1080 Ti, is connected directly to the CPU, rather than using RPC over a network for communication.
Therefore, these results cannot be compared directly to either the \remote or \onprem Brainwave performance; however, they provide a useful characterization of limiting performance.
The purple GPU points utilize the Brainwave implementation of \resnet where, as with the Brainwave implementation on CPU, a \protobuf file is imported.
This is what we would expect within \cmssw for custom models in the future and represents the closest direct comparison of a GPU with the Brainwave FPGA implementation.  
The other GPU lines consist of the official \resnet as provided within \tensorflow. The official \resnet can have better inference times by factors of a few. An optimized version of \resnet is also available. It gives a 0--20\% reduction in inference with respect to the official \resnet.
All of the GPU benchmarks also follow the expected trend for large image batch sizes, with an improvement in the aggregate performance. The per-image latency for a batch of one image is found to be anywhere from 5 to 10 times worse than the ultimate performance on a GPU. 
\end{sloppypar}

Within \cmssw, we find that importing the \protobuf model of \resnet can take approximately 5 minutes. Once the model is imported, subsequent inferences take, on average, 1.75~seconds per inference.  
This benchmark point can most closely be compared with the standalone single-thread CPU performance that is shown in Fig.~\ref{fig:cpu}, approximately 500~ms.
The main differences between the standalone performance and the \cmssw tests are two-fold: the \tensorflow version (1.06 vs. 1.10) and the processor speed (2.6~GHz vs. 3.6~GHz).
It is not uncommon for hardware across the global computing grid of the CMS experiment to vary in performance significantly, which is another consideration when deploying both \onprem and \remote services.

To summarize, for total inference time for a batch of one image, we present Brainwave, CPU, and GPU performance in Table~\ref{tab:comp}.
The most straightforward comparison with the current CMSSW performance of 1.75~seconds is the 10 (60)~ms \onprem (\remote) that it would take to perform inference with Brainwave.
This represents a factor of 175 (30) speedup for Brainwave \onprem (\remote) over current \cmssw CPU performance.  
We can extrapolate from Table~\ref{tab:comp} that, for more modern versions of \tensorflow and CPUs, the \cmssw CPU inference time could improve to approximately 500~ms.

\begin{table*}[t]
\centering
\caption{A summary comparison of total inference time for Brainwave, CPU, and GPU performance
  \label{tab:comp}}
  \begin{tabular}{ccccc} 
    \textbf{Type} & \textbf{Hardware} & \textbf{$\langle$Inference time$\rangle$} & \textbf{Max throughput} & \textbf{Setup}\\
    \hline
    CPU & Xeon 2.6 GHz, 1 core & 1.75 seconds & 0.6 img/s& \cmssw, {\tt TF v1.06} \\
    CPU & i7 3.6 GHz, 1 core & 500~ms & 2 img/s &   python, {\tt TF v1.10} \\
    CPU & i7 3.6 GHz, 8 core & 200~ms & 5 img/s &   python, {\tt TF v1.10} \vspace{1ex}\\
    GPU (batch=1) & NVidia GTX 1080 & 100~ms   & 10 img/s &   python, {\tt TF v1.10} \\ 
    GPU (batch=32)  & NVidia GTX 1080 & 9~ms     & 111 img/s &   python, {\tt TF v1.10} \\ 
    GPU (batch=1) & NVidia GTX 1080 & 7~ms     & 143 img/s &  TF internal, {\tt TF v1.10} \\ 
    GPU (batch=32)  & NVidia GTX 1080 & 1.5~ms   & 667 img/s &  TF internal, {\tt TF v1.10} \vspace{1ex}\\
    Brainwave & Altera Artix & 10~ms & 660 img/s &  \cmssw, \onprem \\
    Brainwave & Altera Artix & 60~ms & 660 img/s &  \cmssw, \remote  
\end{tabular}
\end{table*}

GPU comparisons can be more nuanced\footnote{For that matter, CPU comparisons can also be nuanced when considering devices with many cores and large RAM. However, they do not fit in with the \cmssw computing model.},
depending on the model implementation and batch sizes.  
However, for a batch of one image, we can say that the Brainwave inference latencies, both \onprem and \remote including network latencies, are of a similar order to local, physically connected GPU inference times.
The GPU and Brainwave have similar maximum throughput, about 660 images per second, though the former only achieves this with large batch size and the latter achieves this when accessed with many CPUs simultaneously.  
It should be emphasized that Brainwave achieves this performance using single-image requests and including network infrastructure for deployment as a service, while the GPU requires a large batch size for the same performance and is directly connected to the CPU via PCIe (Peripheral Component Interconnect express).
As will be described in Sec.~\ref{sec:outlook}, future studies are needed to better understand the scalability and cost of different heterogeneous computing architectures.  The performance of other coprocessors as services, including GPUs, is another item for future study.

\section{Summary and outlook}
\label{sec:outlook}
The current computing model for particle physics will not suffice to keep up with the expected future increases in dataset size, detector complexity, and event multiplicity.
Single-threaded CPU performance has stagnated in recent years; therefore, it is no longer viable to rely on improvements in the clock speed of general-purpose computing.
Industry trends towards heterogeneous computing---mixed hardware computing platforms with CPUs communicating with GPUs, FPGAs, and ASICs as coprocessors---provide a potential solution that can perform calculations more than an order of magnitude faster than CPUs.
The new coprocessor hardware is geared towards machine learning algorithms, which are parallelizable, high-performing even with reduced precision, and energy efficient.  
Therefore, to best utilize the new computing hardware, it is important to adopt machine learning algorithms in particle physics computing.  
Fortunately, machine learning is very common in particle physics, from simulation to reconstruction and analysis, and its usage continues to grow.

\begin{sloppypar}In this paper, we explore the potential of FPGAs to accelerate machine learning inference for particle physics computing.
We focus on the acceleration of the \resnet convolutional neural network model and adapt it to physics applications.
As an example, we interpret jets, collimated sprays of particles produced in LHC collisions, as 2D images that are classified by \resnet.
We keep the same architecture but train new weights to distinguish top quark jets from light quark and gluon jets.
Using a publicly available dataset, we compare our model against other state-of-the-art models in the literature and find similarly excellent performance.
We also discuss the potential for Brainwave to be used in other particle physics applications.
For example, neutrino event reconstruction deploys large convolution neural networks in their experiments and large network inferences are a bottleneck in their current computing workflow.  
Coprocessor-accelerated machine learning inference could be deployed for such neutrino experiments {\it today}.
\end{sloppypar}

We accelerate \resnet using the newly available Microsoft Brainwave platform that deploys FPGA coprocessors {\it as a service}.  
We find that using machine learning acceleration {\it as a service} is a simple yet very high-performing approach that can be integrated into modern particle physics experimental software with little disruption.
Using open source RPC protocols, we can communicate with Brainwave from our datacenters with our experimental software to accelerate machine learning inference.
We refer to this workflow as \sonic (Services for Optimized Network Inference on Coprocessors).

Even including the network transit time from the Fermilab datacenter in Illinois to the Microsoft datacenter in Virginia, the inference latency is still 30 times faster than our current, default CPU performance.  
We test Brainwave both as a cloud service and an edge (\emph{on-premises}) service with \resnet inferences averaging 60 and 10~ms, respectively.
For the edge scenario including network service infrastructure, this is comparable to the performance of a GPU connected directly to the CPU for a batch of one image, which is important for the particle physics event processing model.
We also study the scalability of the \sonic workflow by having many batch CPU jobs make requests to a single FPGA service.
We find, even in very extreme scenarios where the job's only task is to access the Brainwave service, 50--100 simultaneous CPU jobs can be executed with little drop in latency while greatly improving the throughput of the FPGA to the point where a GPU can only be competitive with large batch sizes.
This result suggests a setup with many CPUs connecting to one service will be more than sufficient for our computing needs and be more cost-effective.

This proof-of-concept work has potentially revolutionary implications for many large scale scientific experiments.
Further academic studies and industry developments will help to bring this technology to maturity; we highlight a few in particular. 
\begin{itemize}
	\item {\it Continue efforts to design machine learning algorithms to replace particle physics algorithms.} New commercial coprocessors are being designed with machine learning applications in mind, and particle physics should capitalize on this.
	\item {\it Develop tools for generically translating models and explore a broad offering of potential hardware.} While we have explored a specific \resnet network architecture, machine learning algorithms for different types of physics applications will require very different network architectures.  We will need to explore all the available tools to automate network translation for specialized hardware.  Various available hardware options coming onto the market should be explored and benchmarked.
	\item {\it Continue to build infrastructure and study scalability/cost.} We have developed a minimal experimental software framework for communicating with Brainwave. This will have to grow in sophistication for authentication, communication, flexibility, and scalability to operate within the worldwide grid computing paradigm.  
\end{itemize}

Future heterogeneous computing architectures are a powerful and exciting solution to particle physics computing challenges.
This study is the first demonstration of how to integrate them into our physics algorithms and our computing model to enable new discoveries in fundamental physics.

\section*{Acknowledgements}
\label{sec:acknowledgements}

\begin{sloppypar}We would like to thank the entire Microsoft Azure Machine Learning, Bing, and Project Brainwave teams for the development of and opportunity to preview and study the acceleration platform. 
In particular, we would like to acknowledge Doug Burger, Eric Chung, Jeremy Fowers, Daniel Lo, Kalin Ovtcharov, and Andrew Putnam, for their support and enthusiasm.
We would like to thank Lothar Bauerdick and Oliver Gutsche for seed funding through USCMS computing operations. 
We would like to thank Alex Himmel and other \nova~collaborators for support and comments on the manuscript.
\end{sloppypar}

Part of this work was conducted at ``\textit{iBanks},'' the AI GPU cluster at
Caltech. We acknowledge NVIDIA, SuperMicro, and the Kavli Foundation
for their support of ``\textit{iBanks}.''
Part of this work was conducted using Google Cloud resources provided by the MIT Quest for Intelligence program.
Part of this work is supported through IRIS-HEP under NSF-grant 1836650.
We thank the organizers of the public available top tagging dataset (and others like it) for providing benchmarks for the physics community. 

The authors thank the {\nova} collaboration for the use of its Monte Carlo
software tools and data and for the review of this manuscript.
This work was supported by the US Department of Energy and the US
National Science Foundation. {\nova} receives additional support from the
Department of Science and Technology, India; the European Research
Council; the MSMT CR, Czech Republic; the RAS, RMES, and RFBR, Russia;
CNPq and FAPEG, Brazil; and the State and University of Minnesota. We
are grateful for the contributions of the staff at the Ash River
Laboratory, Argonne National Laboratory, and Fermilab. 

On behalf of all authors, the corresponding author states that there is no conflict of interest. 

\bibliographystyle{spphys}       
\bibliography{aml}


\end{document}